%% file: main.tex
\documentclass{article}
\usepackage{graphicx,array,hyperref,setspace,biblatex}
\usepackage{amsmath,amsthm,amssymb,amsfonts}
\usepackage[margin=0.8in]{geometry}
\usepackage{authblk}
\usepackage{bm}
\usepackage[linesnumbered,ruled]{algorithm2e}
\usepackage{algpseudocode}
\usepackage{subfig}
\usepackage{enumitem}
\newcommand{\R}{\ensuremath{\mathbb{R}}}
\newcommand{\N}{\ensuremath{\mathbb{N}}}
\newcommand{\B}{\ensuremath{\mathbb{B}}}

\title{Comparing Quantum Annealing and Spiking Neuromorphic Computing for Sampling Binary Sparse Coding QUBO Problems}

\author[1]{Kyle Henke\footnote{These two authors contributed equally to this work.}\thanks{Email: khenke@lanl.gov}}
\author[1]{Elijah Pelofske\textsuperscript{*}\thanks{Email: epelofske@lanl.gov}}
\author[1]{Garrett Kenyon\thanks{Email: gkenyon@lanl.gov}}
\author[2]{Georg Hahn\thanks{Email: ghahn@hsph.harvard.edu}}
\affil[1]{Los Alamos National Laboratory, CCS-3 Information Sciences}
\affil[2]{Harvard University, T.H. Chan School of Public Health}

\date{\vspace{-6ex}}
\begin{document}

\maketitle
\input{abstract}

\input{text}

\section*{Acknowledgments}
This work was supported by the U.S.\ Department of Energy through Los Alamos National Laboratory. Los Alamos National Laboratory is operated by Triad National Security, LLC, for the National Nuclear Security Administration of the U.S.\ Department of Energy (under the contract No.~89233218CNA000001). The research presented in this article was supported by the Laboratory Directed Research and Development program of Los Alamos National Laboratory under the project number~20220656ER and 20240032DR. This research used resources (D-Wave Leap Cloud access) provided by the Los Alamos National Laboratory Institutional Computing Program, which is supported by the U.S. Department of Energy National Nuclear Security Administration under Contract No. 89233218CNA000001. Research presented in this article was supported by the NNSA's Advanced Simulation and Computing Beyond Moore's Law Program at Los Alamos National Laboratory. We gratefully acknowledge support from the Advanced Scientific Computing Research (ASCR) program office in the Department of Energy's (DOE) Office of Science, award \#77902. This research used resources provided by the Darwin testbed at Los Alamos National Laboratory (LANL) which is funded by the Computational Systems and Software Environments subprogram of LANL's Advanced Simulation and Computing program (NNSA/DOE). This paper has been assigned LANL report number LA-UR-23-26361

\section*{Data Availability Statement}
Data is publicly available on Zenodo \cite{pelofske_2025_14735999}.

\setlength\bibitemsep{0pt}
\printbibliography
\end{document}

%% file: abstract.tex
\begin{abstract}
We consider the problem of computing a sparse binary representation of an image. To be precise, given an image and an overcomplete, non-orthonormal basis, we aim to find a sparse binary vector indicating the minimal set of basis vectors that when added together best reconstruct the given input. We formulate this problem with an $L_2$ loss on the reconstruction error, and an $L_0$ (or, equivalently, an $L_1$) loss on the binary vector enforcing sparsity. This yields a quadratic unconstrained binary optimization problem (QUBO), whose optimal solution(s) in general is NP-hard to find. The contribution of this work is twofold. First, we solve the sparse representation QUBOs by solving them both on a D-Wave quantum annealer with Pegasus chip connectivity via minor embedding, as well as on the Intel Loihi~2 spiking neuromorphic processor using a stochastic Non-equilibrium Boltzmann Machine (NEBM). Second, we deploy Quantum Evolution Monte Carlo with Reverse Annealing and iterated warm starting on Loihi~2 to evolve the solution quality from the respective machines. The solutions are benchmarked against simulated annealing, a classical heuristic, and the optimal solutions are computed using CPLEX. Iterated reverse quantum annealing performs similarly to simulated annealing, although simulated annealing is always able to sample the optimal solution whereas quantum annealing was not always able to. The Loihi~2 solutions that are sampled are on average more sparse than the solutions from any of the other methods. We demonstrate that both quantum annealing and neuromorphic computing are suitable for binary sparse coding QUBOs, and that Loihi~2 outperforms a D-Wave quantum annealer standard linear-schedule anneal, while iterated reverse quantum annealing performs much better than both unmodified linear-schedule quantum annealing and iterated warm starting on Loihi~2. 
\end{abstract}

%% file: text.tex
\section{Introduction}
\label{section:introduction}
The computation of a sparse binary reconstruction of an image is of importance whenever a full image cannot be directly observed, in which case it has to be reconstructed from a sample or projection using compressive sensing. Such scenarios occur, amongst others, in radioastronomy, molecular imaging, and image compression \cite{Ting2006, Mohideen2021}.

Mathematically, we are given an overcomplete and non-orthonormal basis of $p \in \N$ vectors $D = \{ D_1,\ldots,D_p \}$, and we aim to infer a sparse representation of a given image $\boldsymbol{x}$. Here, an overcomplete set is defined as one that contains more functions than needed for a basis. All basis vectors and the image $\boldsymbol{x}$ are assumed to be of equal dimension $m \in \N$. The task is to find the minimal set of non-zero activation coefficients $\boldsymbol{a}$ that accurately reconstruct $\boldsymbol{x}$, where $\boldsymbol{a} \in \B^p$ is a binary vector of length $p$ for $\B=\{0,1\}$. We can express the computation of a sparse binary representation of the image $\boldsymbol{x}$ using the basis $D$ as the minimization of the energy function
\begin{equation}
    E \left( \boldsymbol{a} \right) = \min_{\boldsymbol{a}} \left[ \frac{1}{2} \Vert \boldsymbol{x} - D \boldsymbol{a} \Vert_2^2 + \lambda \Vert \boldsymbol{a} \Vert_0 \right],
    \label{eq:objective}
\end{equation}
where $\Vert \cdot \Vert_2$ is the Euclidean norm and $\Vert \cdot \Vert_0$ denotes the number of nonzero elements. The operation $D \boldsymbol{a}$ denotes the addition of all basis matrices in $D$ that correspond to an entry of $1$ in $\boldsymbol{a}$. The parameter $\lambda>0$ is a Lasso-type penalization parameter controlling the sparseness of the solution \cite{Tibshirani1996}. A large value of $\lambda$ results in a more sparse solution to eq.~\eqref{eq:objective}, while smaller values yield denser solutions. Therefore, the parameter $\lambda$ allows one to effectively balance the reconstruction error (the $L_2$ norm) and the number of non-zero activation coefficients (the $L_0$ norm). Since eq.~\eqref{eq:objective} belongs to the class of 0-1 integer programming problems, finding a sparse representation falls into an NP-hard complexity class. The energy function of eq.~\eqref{eq:objective} is non-convex and typically contains multiple local minima. 

We investigate two types of computing models to solve the sparse representation problem given in eq.~\eqref{eq:objective}. The first is a D-Wave quantum annealer with chip ID \texttt{Advantage\_system4.1}. Moreover, we employ the spiking neuromorphic computing processor Loihi~2 by Intel. Quantum annealing is a type of analog quantum computation that aims to use quantum fluctuations in order to search the objective function landscape of a combinatorial optimization problem for the global minimum \cite{Kadowaki_1998, Farhi2000}. Quantum annealer devices are manufactured by D-Wave, which use superconducting flux qubits to implement quantum annealing with a Transverse field Hamiltonian and have been studied for sampling a variety of problem types \cite{Johnson2011, Lanting_2014}. The Transverse field driving Hamiltonian facilitates state transitions. Although current quantum annealers are subject to various sources of errors, limited coherence times, and limited hardware connectivity, there is evidence that quantum annealing can provide an advantage over classical algorithms for certain types of problems \cite{King2021scaling, Tasseff2022, king2024computational, bauza2024scaling}. Neuromorphic computing is a proposed computing model inspired by the human brain, which is able to complete certain learning tasks better than classical von Neumann computers \cite{Davies2021, Schuman2022}. To the best of our knowledge, this is the first demonstration of an exact $L_0$ constraint problem being solved on two the two computing paradigms, respectively.

This article is a significant extension of \cite{henke2023sampling}, in which the unsupervised dictionary learning approach to generate sparse coding QUBO models was introduced for the purposes of sampling those problems on Loihi~1. Among others, this article investigates post-hoc normalization techniques to speed up computations on Loihi~1, optimal parameter selection, reconstruction errors, and unsupervised dictionary learning for Loihi~1 approaches and their classical counterparts. The article is structured as follows. After a literature review in Section~\ref{section:introduction_literature_review}, we introduce the quantum annealing and neuromorphic computing technology we investigate in Section~\ref{section:methods}. Experimental results are presented in Section~\ref{section:results}. The article concludes with a discussion in Section~\ref{section:discussion}. Data generated by this study is publicly available \cite{pelofske_2025_14735999}.

%%%%%%%%%%%%%%%%%%%%%%%%%%%%%%%%%%%%%%%%%%%%%%%%%%%%%%%%%%%%%%
%%%%%%%%%%%%%%%%%%%%%%%%%%%%%%%%%%%%%%%%%%%%%%%%%%%%%%%%%%%%%%
%%%%%%%%%%%%%%%%%%%%%%%%%%%%%%%%%%%%%%%%%%%%%%%%%%%%%%%%%%%%%%
\subsection{Literature Review}
\label{section:introduction_literature_review}
The goal of any sparse coding method is to accurately recover an unknown sparse vector from a few noisy linear measurements. Unfortunately, this estimation problem is NP-hard in general (when formulated as 0-1 integer optimization), and it is therefore usually approached with an approximation method, such as some (C)LASSO (Constrained Least Absolute Shrinkage and Selection Operator) or orthogonal matching pursuit (OMP) \cite{4385788} where the penalty term in eq.~\eqref{eq:objective} is an $L_{1}$ ($\left\| \cdot \right\|_{1}$) constraint. Because the coefficients are allowed to be any positive real number, the problem is turned convex, and a global minimum is achieved at the cost of trading off accuracy for less computational complexity.

A comparison of quantum computing (using a D-Wave quantum annealer) and neuromorphic computing (using the Intel Loihi spiking processor) has been investigated in \cite{Henke2020machine, Henke2020alien}. The binary results of the D-Wave quantum annealer and the rate coded solutions motivated by \cite{Rozell2008} to the $L_1$ problem from Loihi~1 were examined with respect to various metrics (power consumption, reconstruction and classification accuracy, etc), which generally finds both the quantum annealing and neuromorphic processors to be reasonably effective at solving the respective underlying problem -- however, exact state-of-the-art classical algorithms outperform both methods. This line of research is continued in \cite{Henke2022}, who compare sparse coding solutions generated classically by a greedy orthogonal matching pursuit (OMP) algorithm with spike-based solutions obtained with the Loihi~1 neuromorphic processor.

A quantum-inspired algorithm for sparse coding is presented in \cite{Nguyen_2018, Romano2022} by formulating a general sparse coding problem as a quadratic unconstrained binary optimization (QUBO) problem suitable for approximation on current hardware such as D-Wave quantum annealers. The QUBO formulations of \cite{Nguyen_2018, Romano2022} were shown to be competitive with the standard (C)LASSO and OMP methods. Similar research on regularized linear regression for sparse signal reconstruction is presented in \cite{Ide2022}, who also derive a QUBO representation and compare solutions obtained with D-Wave Advantage Systems~4.1 to OMP and conventional (C)LASSO methods, the results for which showed good recovery
success rates using the quantum annealer that is comparable to conventional methods.

Two related problems, binary compressive sensing and binary compressive sensing with matrix uncertainty, are investigated in \cite{Ayanzadeh2019}; this study presents an Ising model formulation that has a ground state that is a sparse solution for the binary compressive sensing problem, though no experimental results on D-Wave devices are presented.

%%%%%%%%%%%%%%%%%%%%%%%%%%%%%%%%%%%%%%%%%%%%%%%%%%%%%%%%%%%%%%
%%%%%%%%%%%%%%%%%%%%%%%%%%%%%%%%%%%%%%%%%%%%%%%%%%%%%%%%%%%%%%
%%%%%%%%%%%%%%%%%%%%%%%%%%%%%%%%%%%%%%%%%%%%%%%%%%%%%%%%%%%%%%
\section{Methods}
\label{section:methods}
In this section we present the hardware being used in this research. After some mathematical considerations on the problem under investigation (Section~\ref{section:methods_transformation}), we give details on Intel's Loihi~2 neuromorphic chip (Section~\ref{section:methods_loihi_implementation}), followed by some implementation details for the D-Wave quantum annealer (Section~\ref{section:methods_Quantum_Annealing}). Section~\ref{section:methods_simulated_annealing} describes the classical simulated annealing implementation which we will compare against. Section~\ref{section:methods_cplex} discusses the exact solution obtained with IBM's CPLEX solver \cite{CPLEX}, which is exact and deterministic.

\subsection{Binary Sparse Coding QUBOs}
\label{section:methods_transformation}
For both Intel's Loihi~2 neuromorphic chip, as well as the D-Wave quantum annealer, the problem being solved has to be given as a QUBO problem. In this formulation, the observable states of any variable are $0$ and $1$. During the solution process, both technologies can be described as effectively treating each spin/neuron in some probabilistic combination of $0$ and $1$, in which the variable is both active and non-active at the same time. After the quantum annealing process or the firing process in the neural net is complete, each variable will return to one of the classical states $0$ or $1$.

We start by reformulating eq.~\eqref{eq:objective} in QUBO form. First we expand out the objective function,
\begin{align*}
    E(\boldsymbol{a}) &= \frac{1}{2} \Vert \boldsymbol{x} - D \boldsymbol{a} \Vert_2^2 + \lambda \Vert \boldsymbol{a} \Vert_0\\    
    &= \frac{1}{2} \left[(\boldsymbol{x} - D \boldsymbol{a})^\top (\boldsymbol{x} - D \boldsymbol{a})\right] + \lambda \sum_{i=1}^p a_i\\
    &= \frac{1}{2} (\boldsymbol{x}^\top \boldsymbol{x} - 2\boldsymbol{x}^\top D \boldsymbol{a} +  \boldsymbol{a}^\top D^\top D \boldsymbol{a}) + \lambda \sum_{i=1}^p a_i
\end{align*}
where $\boldsymbol{a} = (a_1,\ldots,a_p)$. Next, we take the derivative with respect to one component of $\boldsymbol{a}$,

\begin{align}\label{eq:sparse_coding_derivative1}
    \begin{split}
        \partial_{a_{i}}E(\boldsymbol{a})={}& \partial_{a_{i}}\left[\frac{1}{2} (\sum_{j=1}^{m} x_{j}^{2} -2 \sum_{j=1}^{m} ( x_{j} \sum_{i=1}^{p}D_{ji}a_{i}) + \sum_{j=1}^{m} ( \sum_{i=1}^{p}D_{ji}a_{i})^{2})+\lambda \sum_{i=1}^{p}a_{i}\right]
    \end{split} & \notag \\
    \begin{split}
        ={}&  -\sum_{j=1}^{m} x_{j} D_{ji} + \sum_{j=1}^{m} ( \sum_{i=1}^{p}D_{ji}a_{i})D_{ji} +\lambda
    \end{split}  & \notag \\
    \begin{split}
        ={}&  -\boldsymbol{x}^\top D_i+ D_i^\top D\boldsymbol{a}+\lambda. 
    \end{split} 
\end{align}
As expected, multiplying out eq.~\eqref{eq:objective} yields a quadratic form in $\boldsymbol{a}$, and taking the derivative yields an expression in terms of $\boldsymbol{a}$. Hence we can recast our objective function as a QUBO problem. For this we define the following two transformations:

\begin{align}
    \label{eq:hQ1}
    h_i &= -\boldsymbol{x}^\top D_i +  D_i^\top D_i+ \lambda,\\
    Q_i &= D_i^\top D.
    \label{eq:hQ2}
\end{align}
Using eq.~\eqref{eq:hQ1} and eq.~\eqref{eq:hQ2}, where $D_i^\top D_i$ represents the self interaction term and hence can be absorbed into the linear term, we can rewrite eq.~\eqref{eq:objective} as the following QUBO,
\begin{align}
	H(\boldsymbol{h}, Q, \boldsymbol{a}) = \sum_{i=1}^p {h_i a_i} + \sum_{i<j} Q_{ij} a_i a_j,
    \label{eq:H}
\end{align}
which is now in suitable form to be solved on either the Loihi~2 neuromorphic chip or the D-Wave quantum annealer.

\begin{figure}[htbp]
    \centering
    \includegraphics[width=0.49\textwidth]{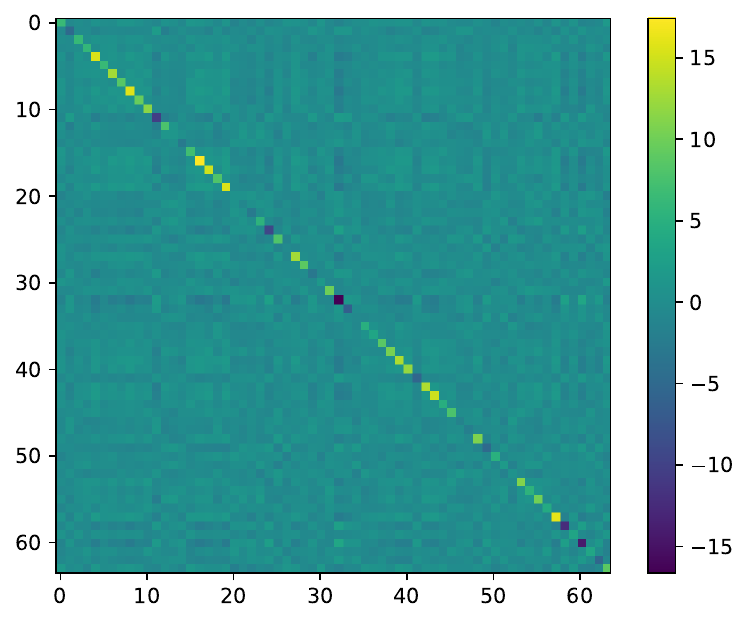}\hfill
    \includegraphics[width=0.49\textwidth]{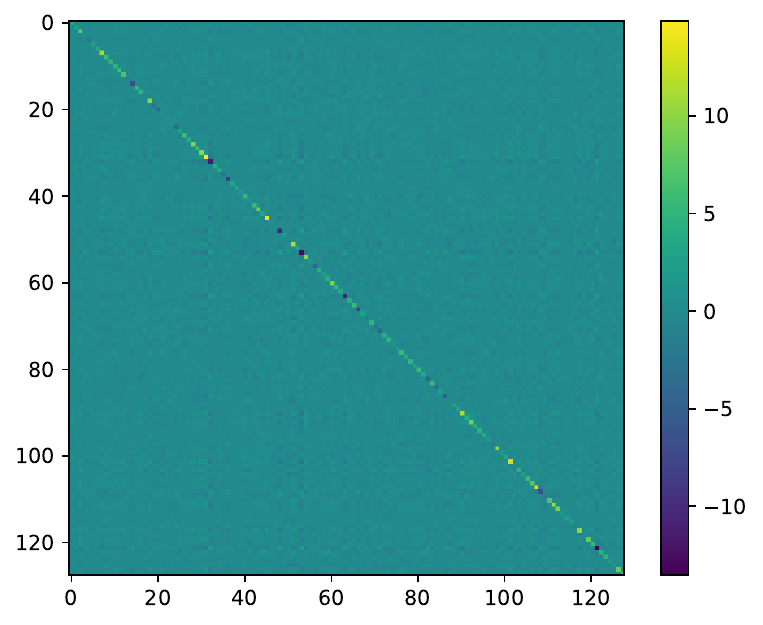}
    \caption{Binary sparse coding QUBO model coefficient matrices. Colormaps of the full QUBO symmetric matrix for a (randomly chosen and representative) sparse coding QUBO with 64 variables (left) and 128 variables (right). Each heatmap corresponds to the coefficient for each of the linear terms (on the diagonal) and for each of the quadratic terms (on the off-diagonals). The axis labels correspond to the matrix row and column indices. }
    \label{fig:QUBO_matrices}
\end{figure}

Figure~\ref{fig:QUBO_matrices} shows a visual representation of both a $64$ and $128$ variable sparse coding QUBO as a symmetric square matrix, where the diagonal terms are the linear terms of eq.~\eqref{eq:H}, and the off-diagonal terms are the quadratic variable interactions. The coefficients for each term are encoded using a colormap. Figure~\ref{fig:QUBO_matrices} shows that the linear terms are typically strongly positively or negatively weighted compared to the off diagonal terms.

The binary sparse coding QUBOs are constructed using the un-normalized dictionary learning introduced in \cite{henke2023sampling}. In particular, the dictionary learning is performed using the classical heuristic simulated annealing using the settings that will be described in Section~\ref{section:methods_simulated_annealing}. As done previously \cite{henke2023sampling}, a single image from the standard fashion MNIST (fMNIST) data set \cite{xiao2017fashionmnist} is used, where the 28x28 image is divided into 16 7x7 patches, and then a dictionary size of $64$ variables is used for each of the $16$ patches. We also evaluate a version where we increase the dictionary size to $128$ variables.

\subsection{Unsupervised Dictionary Learning}
Sparse coding optimization can be viewed as a two-step process. First, a dictionary of basis vectors is learned in an unsupervised manner using a local Hebbian rule. When solving the convex (C)LASSO problem, it is typically necessary to re-normalize the columns of the dictionary \(D\) after each learning epoch. This normalization is essential for convergence in (C)LASSO because the entries of the sparse vector \(a\) can take on any real value.

Here, we use a learning procedure presented in \cite{henke2023sampling} that enables the algorithm to determine the optimal feature norms based on a specified target average sparsity level, \(s \in (0, 1)\). The dictionary is initialized with features drawn from a normal distribution, ensuring their norms are randomly below 1, along with a small sparsity penalty parameter \(\lambda\). After solving the binary sparse coding problem for each sample in the training set, the dictionary is updated. If the average sparsity across the training epoch exceeds the target \(s\), the penalty parameter \(\lambda\) is increased for the next epoch until the average reconstruction error and sparsity converge.

\subsection{Loihi~2 Neuromorphic Chip Implementation}
\label{section:methods_loihi_implementation_Loihi_chip_implementation}

The Intel Loihi 2 neuromorphic processor is designed to replicate core elements of biological neural structure and processing, thereby targeting specific classes of computational tasks. Compared to its predecessor, Loihi 1, Loihi 2 offers several enhancements, including improved manufacturing processes, more connectivity, faster processing, and lower energy consumption. Each Loihi 2 chip contains roughly one million neurons and approximately 120 million synapses.

Loihi hardware employs binary spike events, wherein neurons communicate through digital pulses that can be either “on” or “off” (1 or 0). This binary scheme mirrors the all-or-nothing firing patterns observed in biological neurons, though with lower granularity.

Whereas the Loihi 1 architecture has been utilized with the Stochastic Constraint Integrate and Fire (SCIF) neuron model \cite{Aimone_2022, henke2023sampling} to solve QUBO problems, the expanded capabilities of Loihi 2 support more sophisticated models. One such model is the Non-equilibrium Boltzmann Machine (NEBM) neuron model, available in Intel’s \emph{lava} software package and specifically designed for optimization tasks such as Quadratic Unconstrained Binary Optimization (QUBO). Drawing on principles of statistical mechanics and thermodynamics—particularly the concept of a non-equilibrium steady state \cite{PhysRevLett.112.180602}—the NEBM model departs from traditional approaches by operating under continuously driven conditions rather than settling into thermal equilibrium.

In contrast to equilibrium-based Boltzmann machines, which treat neurons as binary units obeying a Boltzmann distribution, NEBM introduces dynamics influenced by external driving forces. This design encourages more extensive exploration of the solution space and offers a mechanism to escape local minima. Additionally, NEBM does not conserve energy in the traditional sense; it instead dissipates energy due to the open, driven nature of the system. This dissipation is essential for creating an effective annealing process, wherein the system is gradually “cooled” to stabilize into low-energy configurations that correspond to high-quality solutions. Unlike strictly equilibrium-based models, NEBM inherently incorporates time-dependent behavior, allowing for dynamic update rules that can be tuned for faster convergence or more thorough exploration.

When applying NEBM to a QUBO problem, each variable in the QUBO formulation is mapped onto a corresponding neuron, and the pairwise interactions defined by the QUBO matrix translate directly into synaptic connections. In a manner analogous to simulated annealing, NEBM orchestrates a progressive reduction of external driving influences, guiding the network toward low-energy states. The non-equilibrium nature of the model ensures that the system remains driven by both internal neuronal processes and external inputs throughout its evolution. Ultimately, this paradigm is well-suited for deployment on neuromorphic platforms such as Loihi 2, where spiking neural networks can efficiently implement these dynamic, non-equilibrium processes to solve complex optimization problems.

\subsubsection{State Variables and Parameters}
Each NEBM neuron \(i\) maintains at least the following state variables:

\begin{itemize}[noitemsep]
    \item \(\displaystyle v_i(t)\): The membrane potential at time step \(t\).
    \item \(\displaystyle a_i(t)\): The binary \(\{0,1\}\) spike output at time \(t\). 
    \item \(\displaystyle T(t)\): The system temperature (or inverse temperature \(\beta(t)\equiv 1/T(t)\)) follows an annealing schedule.
    \item \(\displaystyle r_i(t)\): A refractory state or counter to modulate firing behavior over time.
    \item \(\displaystyle \alpha, \rho, \kappa, \gamma, \theta\): Hyperparameters controlling decay factors, refractory penalties, noise scales, and thresholds.
\end{itemize}

\subsubsection{Membrane Potential Update}
The discrete-time update for the membrane potential is:
\begin{equation}
v_i(t+1) 
\;=\; \alpha \, v_i(t)
\;+\;\sum_{j} Q_{ij} \, a_j(t)
\;+\; h_i
\;+\; \gamma \,\eta_i\bigl(t; T(t)\bigr),
\label{eq:vm-update}
\end{equation}
where:
\begin{itemize}[noitemsep]
    \item \(\alpha \in [0,1)\) is a decay constant ensuring \(v_i\) does not grow unbounded.
    \item \(Q_{ij}\) is the weight from neuron \(j\) to \(i\), derived from the QUBO matrix \(Q\).
    \item \(h_i\) is the bias term for neuron \(i\), from the vector \(\mathbf{h}\).
    \item \(\gamma\,\eta_i(t; T)\) represents noise injection of a particular distribution, scaled by \(\gamma\). The temperature \(T(t)\) typically controls the noise variance or amplitude.
\end{itemize}

\subsubsection{Spiking / Stochastic ``Boltzmann-like'' Activation}
After updating the membrane potential, the neuron generates an output spike \(s_i(t+1)\).
We denote \(v_i(t+1)\) as the updated membrane potential for neuron \(i\) at time \(t+1\).
Although a deterministic threshold is available, we chose a biologically inspired and exploration-friendly approach to compute the spike output stochastically:
\begin{equation}
\mathbb{P}\bigl(a_i(t+1) = 1\bigr)
\;=\;
\sigma\!\Bigl(\frac{v_i(t+1)}{T(t)}\Bigr)
\;=\;
\frac{1}{1 + e^{-\frac{v_i(t+1)}{T(t)}}},
\label{eq:stoch-activation}
\end{equation}
where \(\sigma(\cdot)\) is the logistic function, and \(T(t)\) is the temperature. The higher the membrane potential (relative to temperature), the greater the probability of spiking.

\subsubsection{Refractory and Reset Dynamics}
A refractory mechanism helps mimic annealing and prevents excessive firing. Our approach is:
\begin{equation}
r_i(t+1) 
\;=\; 
\rho\,r_i(t) \;+\; a_i(t),
\label{eq:refractory-state}
\end{equation}
\begin{equation}
v_i(t+1) 
\;=\; 
\alpha \, v_i(t) 
\;+\;\sum_{j} Q_{ij}\,a_j(t)
\;+\; h_i
\;-\;\kappa\,r_i(t)
\;+\;\gamma\,\eta_i\bigl(t;T(t)\bigr),
\label{eq:vm-refractory}
\end{equation}
where:
\begin{itemize}[noitemsep]
    \item \(\rho \in (0,1)\) is a decay factor for the refractory state \(r_i\).
    \item \(\kappa\) is a refractory scaling term that reduces the membrane potential based on recent spiking.
\end{itemize}
If \(r_i(t)\) exceeds a certain threshold, the neuron can also be prevented from firing.

\subsubsection{Annealing Schedule}
To help the network converge on low-energy solutions, the temperature \(T\) decreases over time:
\begin{equation}
T(t) 
\;=\; 
T_{\max}
\;-\;
\Delta_{T}\,\cdot t,
\label{eq:annealing-schedule}
\end{equation}
where \(T_{\max}\) is the initial temperature and \(\Delta_{T}\) is a constant decrement (for a linear schedule). As \(T\) lowers, the system becomes more likely to settle into low-energy (potentially optimal) configurations rather than continuing to hop among many states.

\subsubsection{Implementation on Loihi}
\label{section:methods_loihi_implementation}
This section guides through several implementation details on Intel's Loihi~2 spiking neuromorphic chip.

The exact hyperparameters available for use to run Loihi~2 are as follows. The neuron model is set to \texttt{nebm-sa-refract}. The annealing schedule is linear. The maximum temperature ($T_{\max}$) is $15$, and the number of steps per temperature is $20$, with temperature delta ($\Delta_{T}$) set to $1$. The parameter \texttt{refract\_scaling} is set to a random integer between 4 and 8, and \texttt{refract} is set to a random integer between $5$ and $10$. The \texttt{refract\_counter} is set to 0. Each simulation is run for a total of 6000 time steps and sampled every 20 time steps, for a total of 300 samples per run.

An additional technique that we introduce is \emph{iterated warm starting} on Loihi~2; we initialize with a standard run, then take the best found solution in that run and initialize the next run with that solution. This is then repeated a total of $100$ times. The initial run is initialized with a random binary vector. As far as we are aware, this is the first description and assessment of using warm starting iteratively on a spiking neuromorphic processor. 

Despite Loihi 2’s advancements, several factors can lead to non-ideal performance when solving QUBO problems with the NEBM model. First, hardware constraints such as limited on-chip memory and communication bandwidth can restrict the size or connectivity of the problem that can be efficiently mapped, which may result in suboptimal partitioning or approximation of the underlying QUBO. Second, the discrete nature of spiking dynamics and the complexity of tuning neuron and synapse parameters mean that finding the right hyperparameters (e.g., annealing schedules, refractory settings) often involves extensive trial and error, leading to potential inefficiencies. Third, real-time spike communication overhead—particularly if signals must be routed across multiple cores or chips—can slow the effective convergence rate. Fourth, noise and variability inherent in spiking hardware may occasionally cause the system to settle in unfavorable local minima, especially for large-scale or high-dimensional problems. Finally, while non-equilibrium approaches can explore more of the solution space, they also rely on carefully calibrated driving forces, and any inaccuracies or mismatches in these driving inputs can impede overall performance. These factors, taken together, highlight that while Loihi 2 shows promise for neuromorphic optimization, further refinements in hardware design, software tooling, and parameter-tuning methodologies are critical to unlocking its full potential.

\subsection{Quantum Annealing Implementation}
\label{section:methods_Quantum_Annealing}

This section gives some background on the D-Wave quantum device we employ to minimize the sparse coding QUBO of eq.~\eqref{eq:H}.

Quantum annealing is a type of analog quantum computation that generally has the goal of computing low energy states in a physical system that correspond to low energy states of a classical spin system (i.e., diagonal Hamiltonian) \cite{Kadowaki_1998, Farhi2000, Santoro2006, morita2008mathematical, das2008colloquium}. A review of quantum annealing can be found in \cite{Hauke_2020}. D-Wave manufactures quantum annealers using superconducting flux qubits \cite{Johnson2011, Lanting_2014}, several generations of which have been utilized in a large number of studies of different optimization problems and physical spin systems \cite{King_2022, harris2018phase, boixo2013experimental, Venturelli_2015, Boixo_2014}. 

The D-Wave quantum annealer is designed to minimize QUBO problems, given by functions of the form of eq.~\eqref{eq:H}. In eq.~\eqref{eq:H}, the linear terms $h_i \in \R$ and the quadratic couplers $Q_{ij} \in \R$ are chosen by the user to define the problem under investigation. The variables $a_1,\ldots,a_p \in \B$ are binary and unknown. The aim is to find a configuration of $a_1,\ldots,a_p$ which minimizes eq.~\eqref{eq:H}. 

Quantum annealing is based on adiabatic quantum computation \cite{Kadowaki_1998, Farhi2000, Santoro2006, morita2008mathematical, das2008colloquium}, where the system is initialized in the ground state of a Hamiltonian that is easy to prepare, in the case of D-Wave this is the Transverse field Hamiltonian. The system is then slowly transitioned to a (diagonal) Hamiltonian of interest, whose ground state is unknown (and, likely hard to find). In ideal adiabatic settings, namely if the transition is sufficiently slow, the system will remain in the ground state, leading to the ground state of the final Hamiltonian. Therefore, this can be used as a method of computing ground states of spin systems (Ising models). The system is described as

\begin{equation}
    H_\text{system}(t) = A(t)H_\text{init} + B(t)H_\text{Ising},
    \label{equation:QA_high_level}
\end{equation}
where $A(t)$ and $B(t)$ define the anneal schedule, $H_\text{init}$ is the Transverse field Hamiltonian, and $H_\text{Ising}$ is the problem Ising model that we wish to find the optimal solution of. Ising models can easily be converted into QUBOs, and vice versa, and since many optimization problems can be formulated as QUBOs, quantum annealing can be used to sample solutions of a large number of optimization problems \cite{Lucas2014, Yarkoni2021}.

The physical spin system that the programmable quantum annealers created by D-Wave implement is defined specifically as

\begin{equation}
    H_\text{system} = - \frac{A(s)}{2} \Big( \sum_i^p \sigma^x_i \Big) + \frac{B(s)}{2} \Big( H_\text{Ising} \Big),
    \label{equation:quantum_annealing_Hamiltonian}
\end{equation}
where $s \in [0,1]$ is a variable called the \emph{anneal fraction} which defines a weighting of both the transverse field and problem Hamiltonians (by $A(s)$ and $B(s)$) at each point in time during the anneal, $\sigma^x_i$ is the Pauli matrix acting on each qubit $i$, and $H_\text{Ising}$ is the problem Ising model that we wish to sample low energy states from. In our case, $H_\text{Ising}$ is the QUBO of eq.~\eqref{eq:H} that has been converted to an Ising model such that all of the variable states are spins. The user can program a variety of additional features on the device, including different anneal schedules and annealing times.

The other D-Wave device feature we use is \emph{reverse quantum annealing}, see Section~\ref{section:methods_Quantum_Annealing_QEMC_iterated_reverse_annealing}. Reverse quantum annealing uses a modified anneal schedule where we initialize the system in a classical spin configuration (each variable is either spin up or spin down). Then, we program an anneal schedule that varies the anneal fraction $s$ such that some proportion of the Transverse field Hamiltonian is introduced into the system, up until the system state is measured (at which point, the anneal fraction must be $s=1$).

The workflow for using the D-Wave quantum annealer is described in the following sections. All quantum annealing results presented in this article were computed on the D-Wave \texttt{Advantage\_system4.1} device. The device-specific values for $A(s)$ and $B(s)$ on \texttt{Advantage\_system4.1} are given in Appendix~\ref{section:appendix_D_Wave_anneal_schedule_calibration}.

%%%%%%%%%%%%%%%%%%%%%%%%%%%%%%%%%%%%%
\subsubsection{Parallel Quantum Annealing Minor Embeddings}
\label{section:methods_Quantum_Annealing_minor_embeddings}
D-Wave \texttt{Advantage\_system4.1} has a logical hardware graph (the connectivity graph of the physical hardware qubits on the QPU) that is called Pegasus \cite{Boothby2020, Zbinden2020, Dattani2019}. Due to manufacturing defects, the chip does not have the full yield of intended hardware qubits and couplers compared to the logical Pegasus graph. The device has $5627$ programmable qubits and $40279$ couplers which allow for tunable interactions between qubits.

Normally, the connectivity graph of the logical qubits in eq.~\eqref{eq:H} does not match the one of the hardware qubits on the D-Wave QPU. In this case, a minor embedding of the problem onto the hardware architecture is necessary. The idea is to represent logical variable states by a path of qubits on the hardware graph (that are connected on the hardware graph), where the couplers between those qubits are bound with ferromagnetic links so that their spin states are incentivized to be the same (specifically, they are penalized via an energy penalty if the spins disagree) \cite{choi2008minor, choi2011minor, PhysRevA.92.042310, Zbinden2020, minorminer, Venturelli_2015}. These paths of physical qubits are called \emph{chains}, and they are intended to encode the logical variable state of one of the variables in the original QUBO problem. 

In order to make more effective use of the large hardware graph of D-Wave's \texttt{Advantage\_system4.1} (compared to the QUBO models which are being embedded), we employ \textit{parallel quantum annealing}, also \emph{tiling}, in which multiple Ising models can be embedded on the chip such that they are disjoint and are solved in parallel during a single anneal \cite{Pelofske2022parallel, Pelofske2022solving, PhysRevA.91.042314}. As shown in Figure~\ref{fig:embeddings_on_pegasus}, we use multiple all-to-all minor embeddings onto the D-Wave hardware graph, thus allowing us to solve the same problem multiple times simultaneously during a single anneal readout cycle. This allows us to obtain more samples per anneal, but also mitigates local biases and errors that may be present on some parts of the chip (as well as differences in the random minor-embeddings). Moreover, parallel quantum annealing partially mitigates the effects of using a single sub-optimal embedding due to multiple different random embeddings being used, and also different parts of the hardware graph being used (since the D-Wave hardware has non-uniform noise properties across the different components). Note that parallel quantum annealing is also referred to as \textit{tiling} in the D-Wave documentation \footnote{\url{https://dwave-systemdocs.readthedocs.io/en/samplers/reference/composites/tiling.html}}. Throughout all experiments, we consider all-to-all embeddings for sizes $64$ and $128$ variables. Those (random) minor embeddings can be computed with the heuristic tool \texttt{minorminer} \cite{minorminer}, applied iteratively to the hardware graph in order to obtain disjoint minor embeddings, using the methods described in \cite{Pelofske2022parallel} which attempts to maximize the number of disjoint minor embeddings that can be embedded on the D-Wave QPU hardware graph.

Figure~\ref{fig:embeddings_on_pegasus} in Appendix~\ref{section:appendix_parallel_minor_embeddings} shows an example of two such embeddings. The first example shown in Figure~\ref{fig:embeddings_on_pegasus} visualizes seven disjoint minor embeddings of size $64$ variables, the second shows two disjoint minor embeddings of size $128$, each time with all-to-all connectivity. Note that these visualizations of the embeddings are intended to show what regions of the chip are being used. In the plots, all edges and nodes which are present in the induced subgraph are being shown, although the actual minor embeddings might have a sparser connectivity structure.

%%%%%%%%%%%%%%%%%%%%%%%%%
\subsubsection{Choice of D-Wave Hardware Parameters}
\label{section:methods_Quantum_Annealing_parameter_choices}
With newer generations of the D-Wave quantum annealers, more and more features have been added which give the user a greater control over the anneal process. The specific parameters being used are listed in this section.

One necessary consequence of the minor embeddings are the presence of chains, that is the representation of a logical qubit as a set of physical hardware qubits on the chip. However, after annealing, it is not guaranteed that all the physical qubits in a chain take the same value (either zero or one), although they technically represent the same logical qubit. Such a chain is called ``broken''. To arrive at a value for the logical qubit in eq.~\eqref{eq:H}, we used the \textit{majority vote} chain break resolution algorithm \cite{Venturelli_2015}.

We employ the D-Wave annealer with an annealing time of $100$ microseconds, and we query $1000$ samples per D-Wave backend call. To compute the chain strength, we employ the uniform torque compensation feature with a UTC prefactor of $0.6$. The UTC computation, given a problem QUBO, attempts to compute a chain strength which will minimize chain breaks while not too greatly disrupting the maximum energy scale programmed on the device \cite{torque}.

For all experiments involving reverse annealing, the schedule used is $\{ [0,1], [10, s], [90, s], [100, 1] \}$, where each tuple defines a point in time (from the start to the end of the anneal process) and an anneal fraction $s$. The anneal fraction is the normalized time used to control how the quantum Hamiltonian is moved from the initial superposition of states to the problem QUBO (which is a classical Hamiltonian) during the anneal \cite{annealing}. The anneal schedule is constructed by linear interpolation between those four points. The reverse anneal schedule we use is symmetric with a pause duration of $80$ microseconds. It has an increasing and decreasing ramp on either side of a pause of duration $10$ microseconds. We vary the anneal fraction $s$ at which the pause occurs.

Moreover, we called the D-Wave quantum annealer with flag \textit{reduce\_intersample\_correlation} enabled for all experiments, which adds a pause in between each anneal so as to reduce self correlations between the anneals (those correlations may exist in time due to the spin bath polarization effect, see \cite{lanting2020probing}). Both parameters \textit{readout\_thermalization} and \textit{programming\_thermalization} were set to $0$ microseconds. The reverse annealing specific parameter \textit{reinitialize\_state} was enabled for all reverse annealing executions, causing the annealer to reapply the initial classical state after each anneal readout cycle \cite{parameters}.

\subsubsection{Quantum Evolution Monte Carlo with Reverse Annealing}
\label{section:methods_Quantum_Annealing_QEMC_iterated_reverse_annealing}
We will use the D-Wave quantum annealer to compute a Monte Carlo chain of reverse quantum anneal sequences, in which the best solution of an initial set of anneal-readout cycles is encoded as the initial state of the next anneal, and then the best solutions found from each iteration are used as the initialized classical spin state of the subsequent (reverse) anneals. This process works on the \textit{logical} problem, meaning after unembedding of all chained qubits.

To be precise, a sequence of reverse anneals are chained together in a Monte Carlo like process, where each subsequent round of reverse anneals is initialized with a classical state that is defined by the best solution found at the last set of reverse anneals. This chain of reverse anneals is initialized with the best solution found from a $100$ microsecond forward anneal with $1000$ samples (qubit measurements) and the standard linear anneal schedule (forward anneal denotes that it is a standard annealing process, initialized in a uniform superposition and not in a specific configuration of spins). Because of the use of parallel quantum annealing, there is a multiplier on the total number of samples obtained at each QEMC step; for the $64$ variable QUBO problems we actually obtain $7,0000$ samples per $1,000$ sample anneal readout cycle, and for the $128$ variable QUBO problems we obtain $2,000$ samples per $1,000$ sample anneal readout cycle. Each reverse annealing step in the chain uses the parameter \texttt{reinitialize\_state=True} when executing on the D-Wave quantum annealers, which re-initializes the state of the reverse anneal after each anneal-readout cycle. This technique has been used in other contexts, and is referred to as both ``iterative reverse annealing'' \cite{Yamashiro2019, Bando2022, Arai2021, pelofske2023initial} and ``quantum evolution Monte Carlo'' (QEMC) \cite{King2021magnetization, King2018, King2021scaling, Kairys2020, LopezBezanilla2023}. The technique we use here differs slightly from some of these approaches in that we do not use a single anneal-readout cycle at each step of the Monte Carlo chain process, instead we use a batch of anneals at each step and initialize the next anneal with the lowest energy solution found in the current batch of anneals.

\begin{figure}[ht!]
    \centering
    \includegraphics[width=0.26\textwidth]{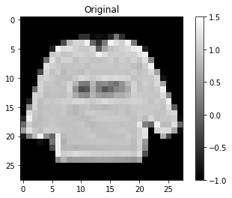}
    \caption{Reference fMNIST \cite{xiao2017fashionmnist} image that will be reconstructed using binary sparse coding. }
    \label{figure:original_image}
\end{figure}

\begin{figure}[ht!]
    \centering
    \includegraphics[width=0.9\textwidth]{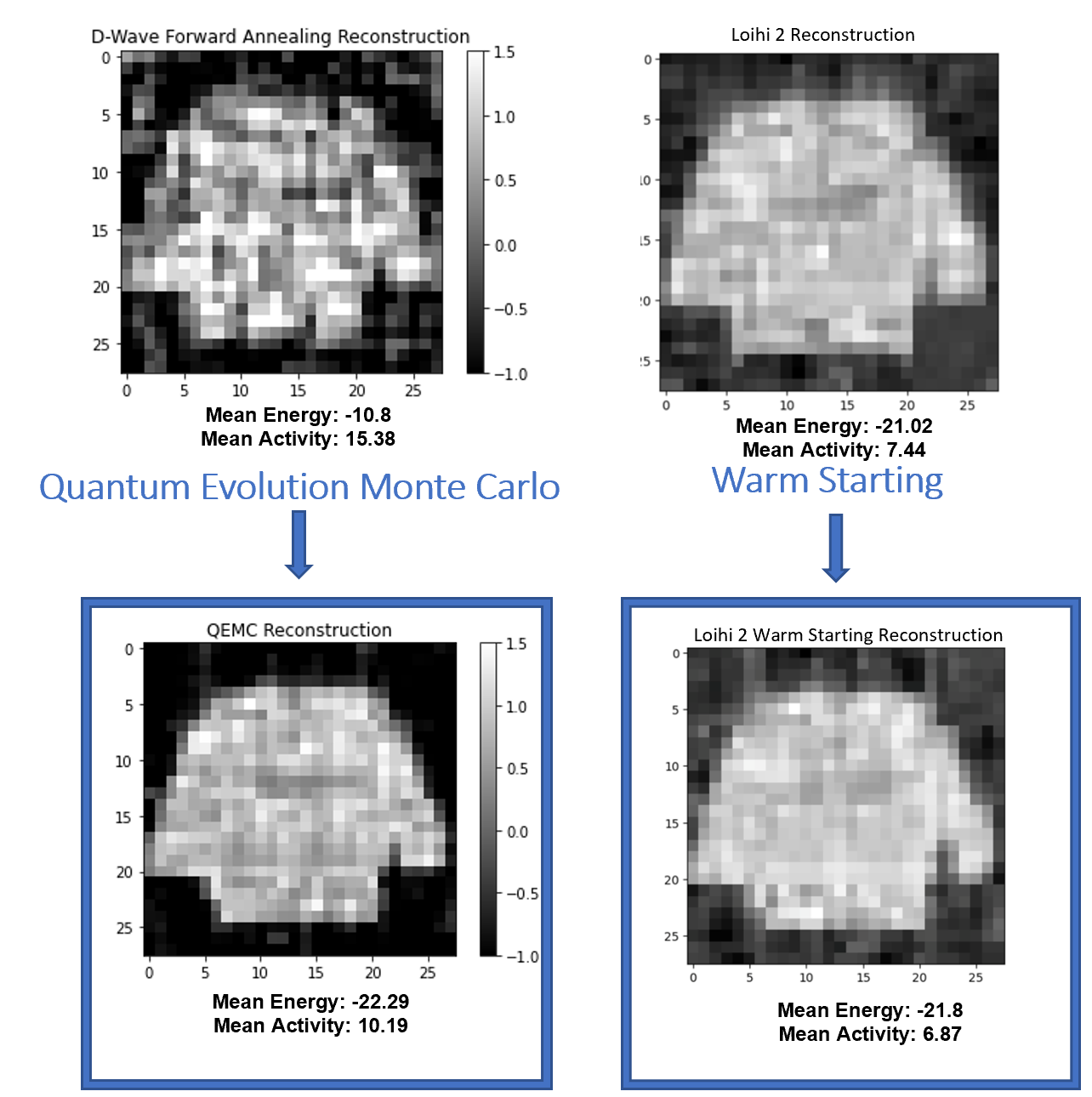}
    \caption{Binary sparse coding image reconstructions of the original image shown in Figure~\ref{figure:original_image}. Top left: The best (lowest energy) reconstruction found using D-Wave with standard forward annealing (with no parallel QA and an annealing time of $20$ microseconds). Bottom left: reverse quantum annealing chain of Monte Carlo-like iterations (QEMC). Top right: Single execution of neuromorphic computing on Loihi~2. Bottom right: Iterated warm starting neuromorphic computing with Loihi~2. Combined, the bottom row of plots show that the two iterative approaches yield better solution quality than that non iterative improvement methods shown in the top row.
    All experimentally computed figure reconstructions are the best mean energy and best mean sparsity across the $16$ QUBO models (for the best parameter combination found for each device or technique).  }
    \label{figure:image_reconstructions}
\end{figure}

\begin{figure}[ht!]
    \centering
    \includegraphics[width=0.24\textwidth]{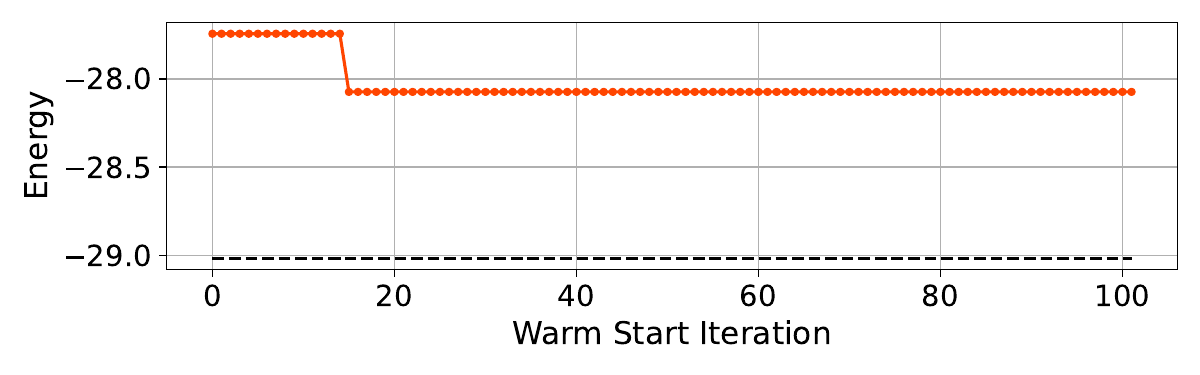}
    \includegraphics[width=0.24\textwidth]{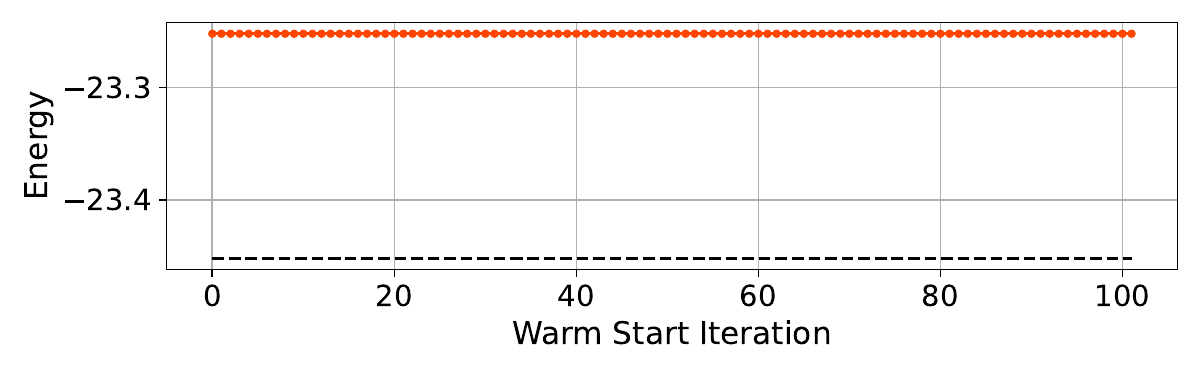}
    \includegraphics[width=0.24\textwidth]{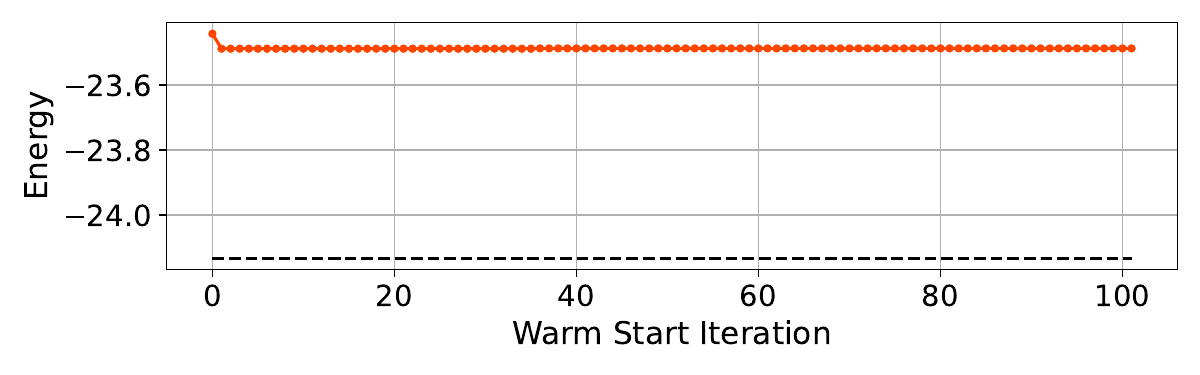}
    \includegraphics[width=0.24\textwidth]{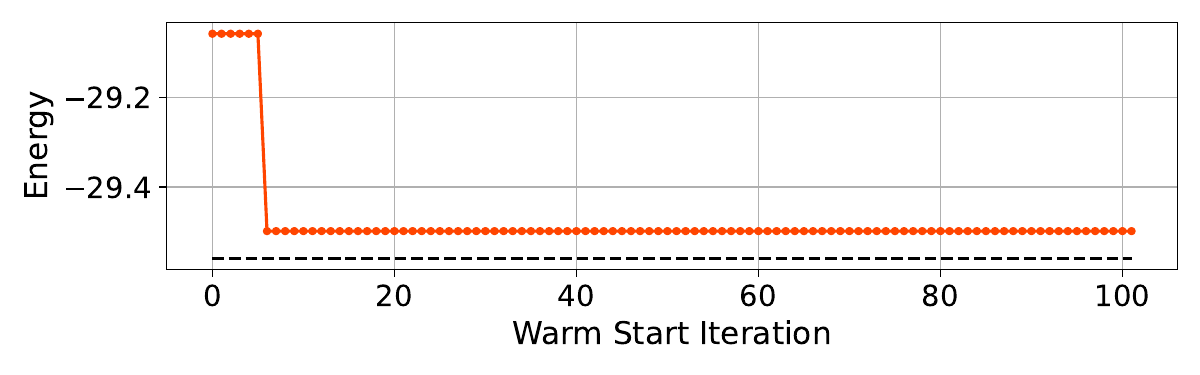}
    \includegraphics[width=0.24\textwidth]{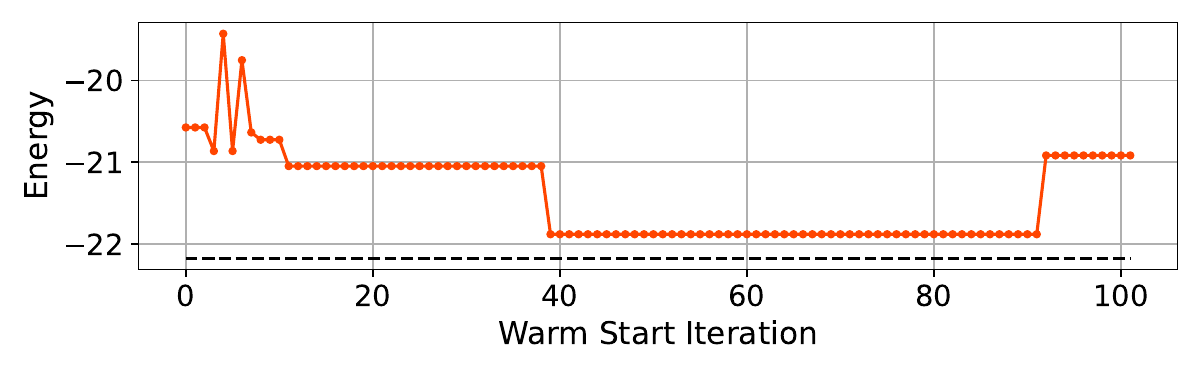}
    \includegraphics[width=0.24\textwidth]{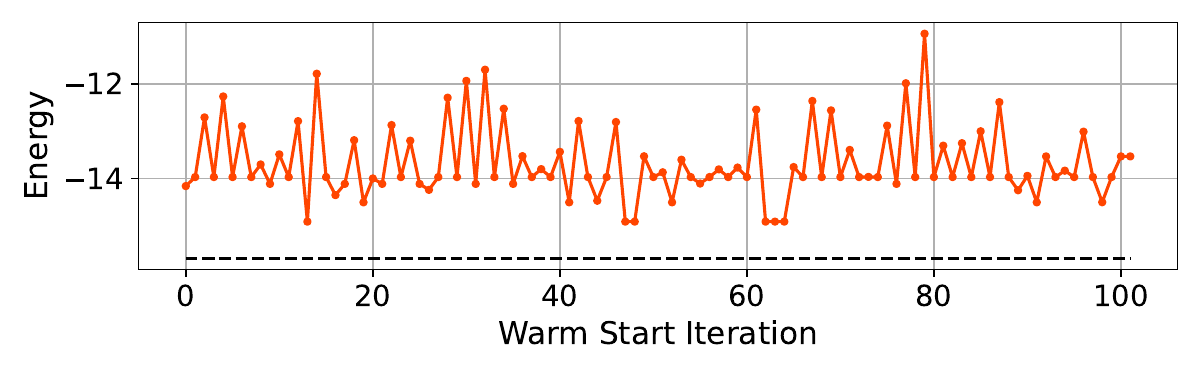}
    \includegraphics[width=0.24\textwidth]{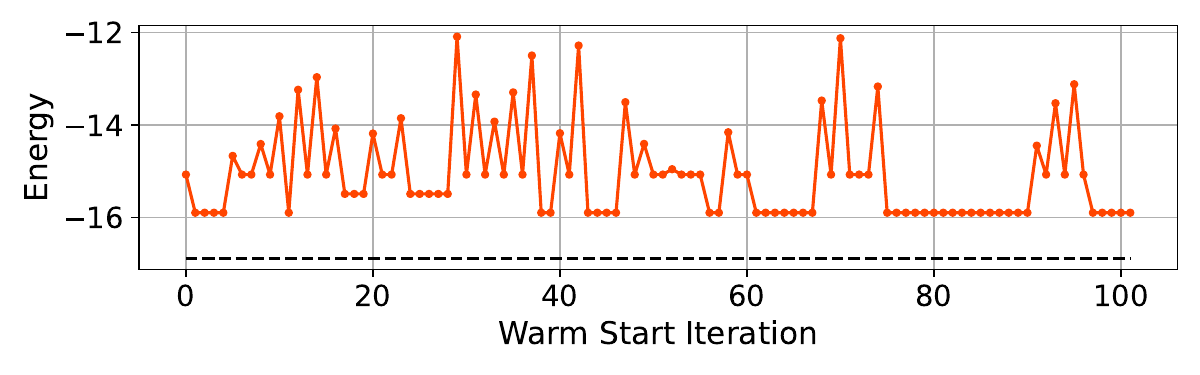}
    \includegraphics[width=0.24\textwidth]{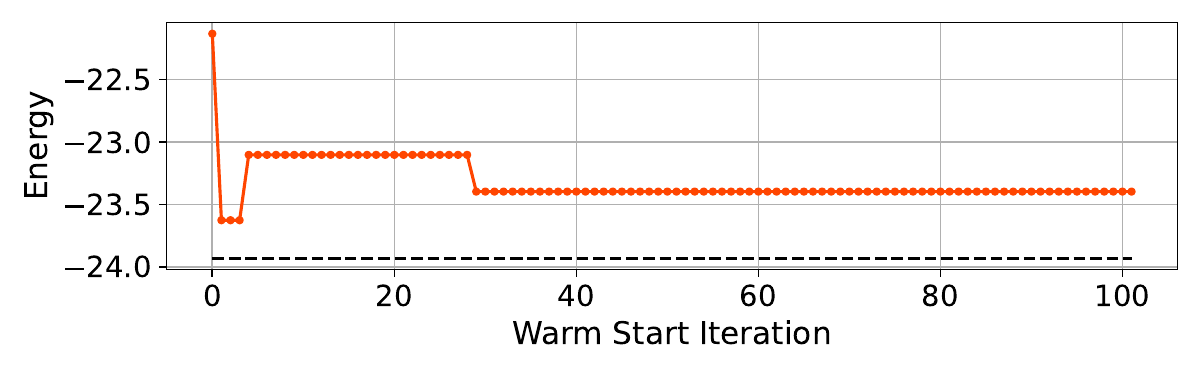}
    \includegraphics[width=0.24\textwidth]{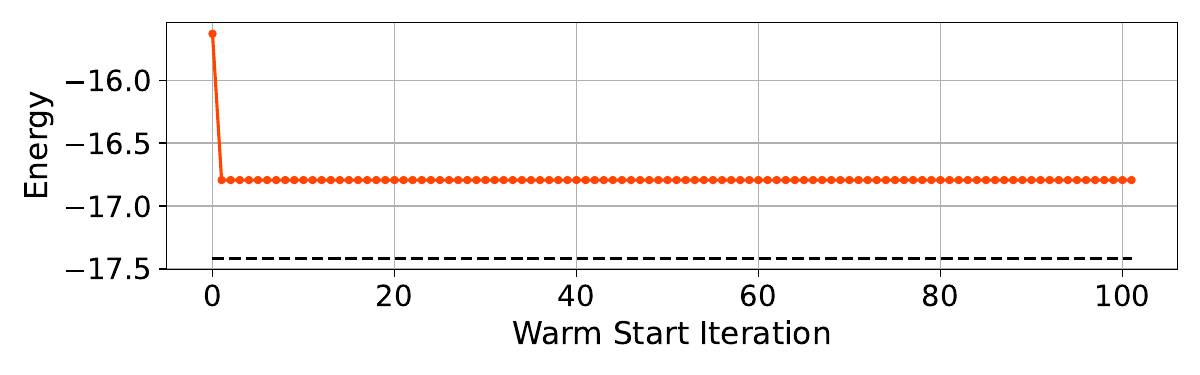}
    \includegraphics[width=0.24\textwidth]{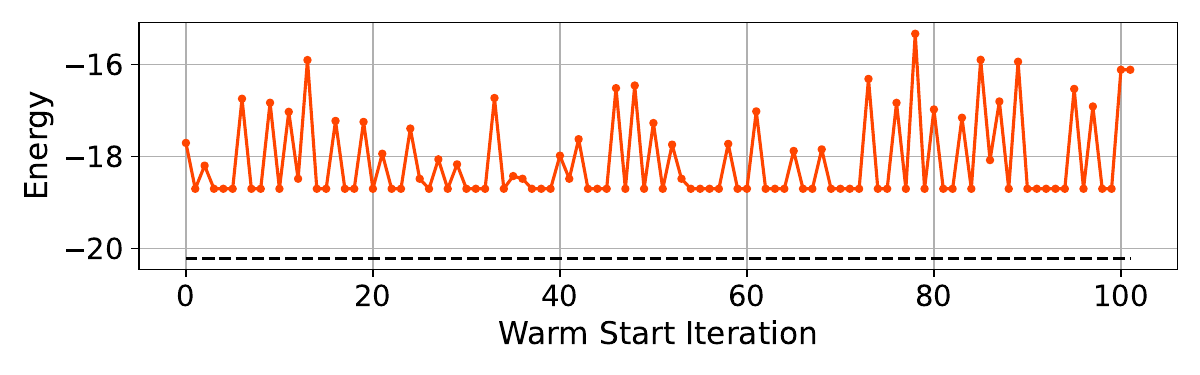}
    \includegraphics[width=0.24\textwidth]{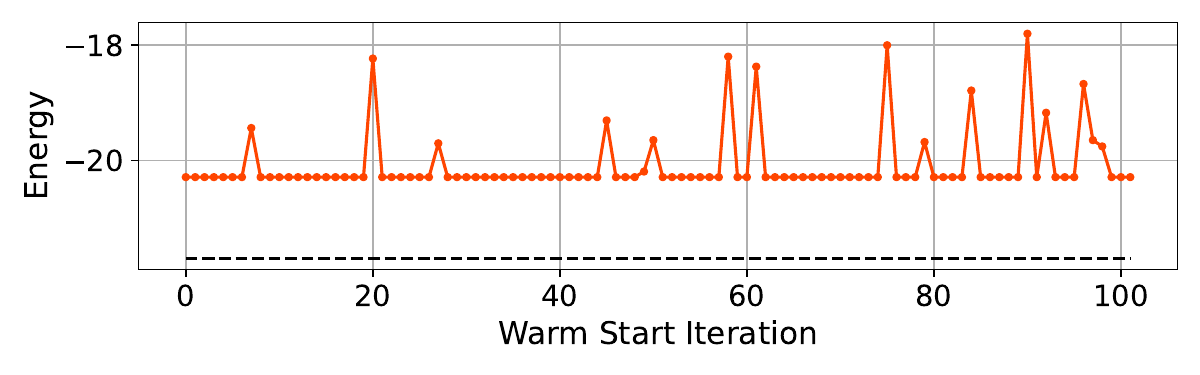}
    \includegraphics[width=0.24\textwidth]{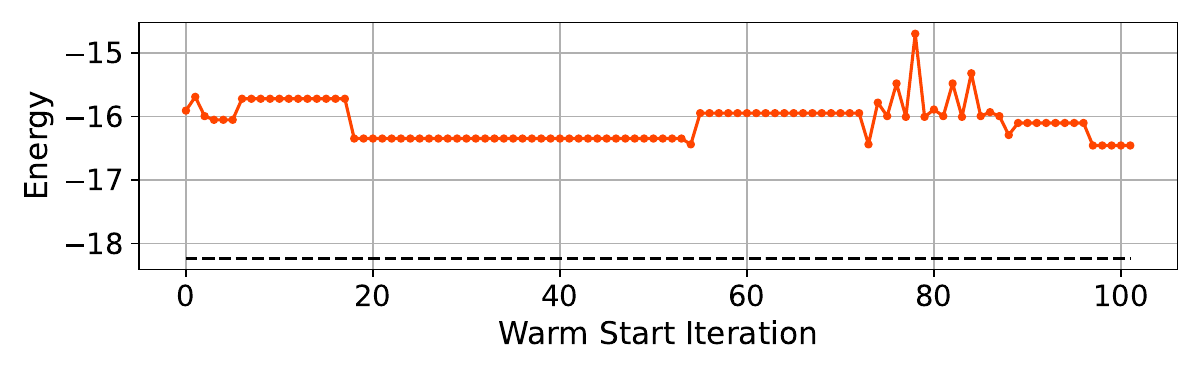}
    \includegraphics[width=0.24\textwidth]{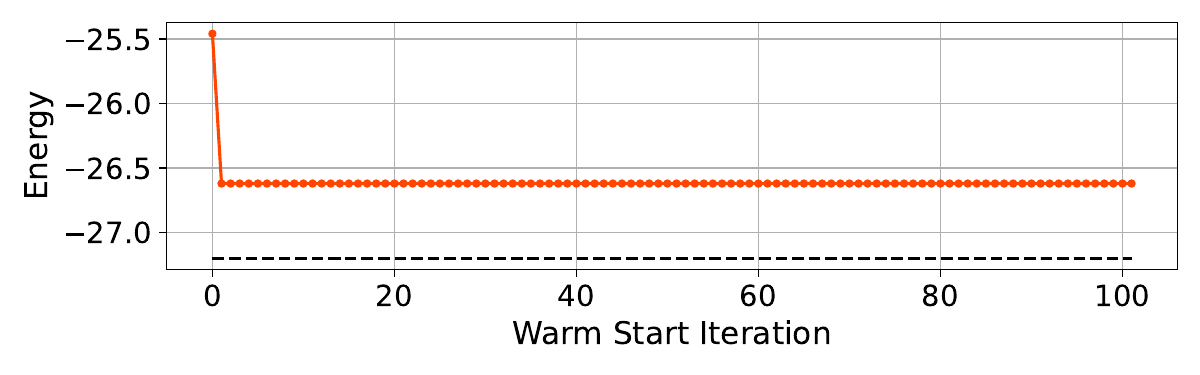}
    \includegraphics[width=0.24\textwidth]{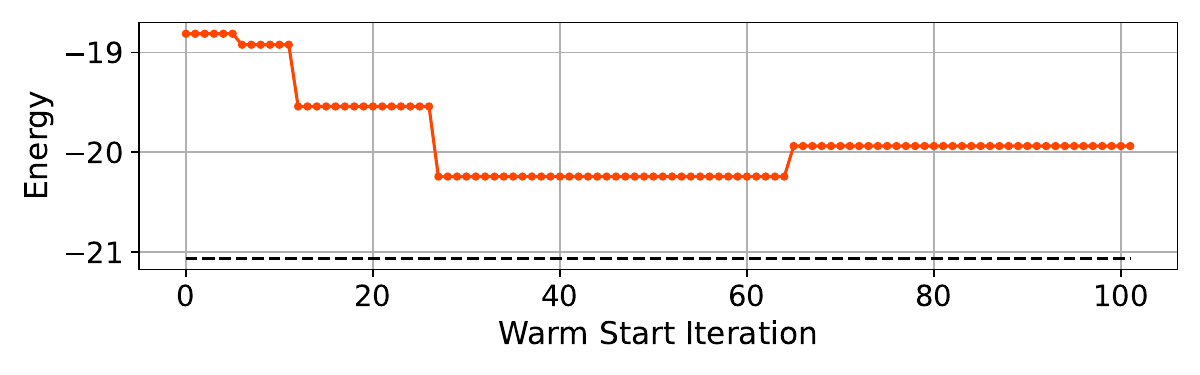}
    \includegraphics[width=0.24\textwidth]{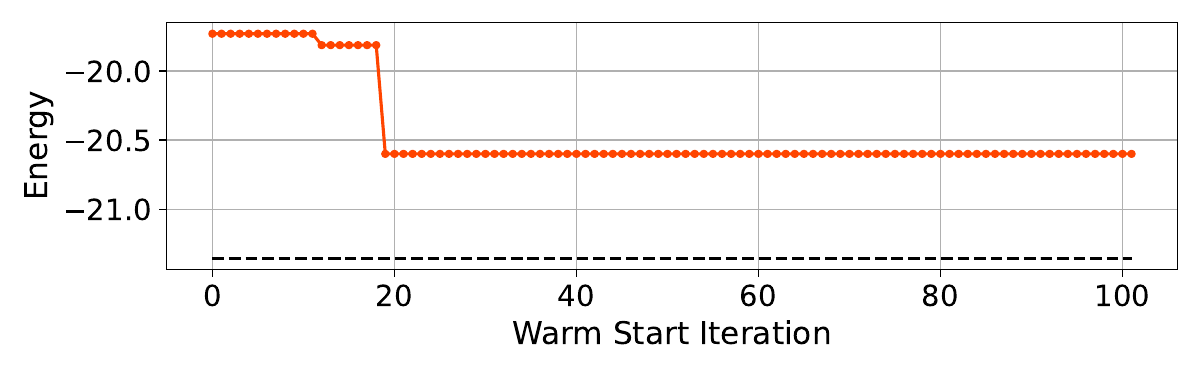}
    \includegraphics[width=0.24\textwidth]{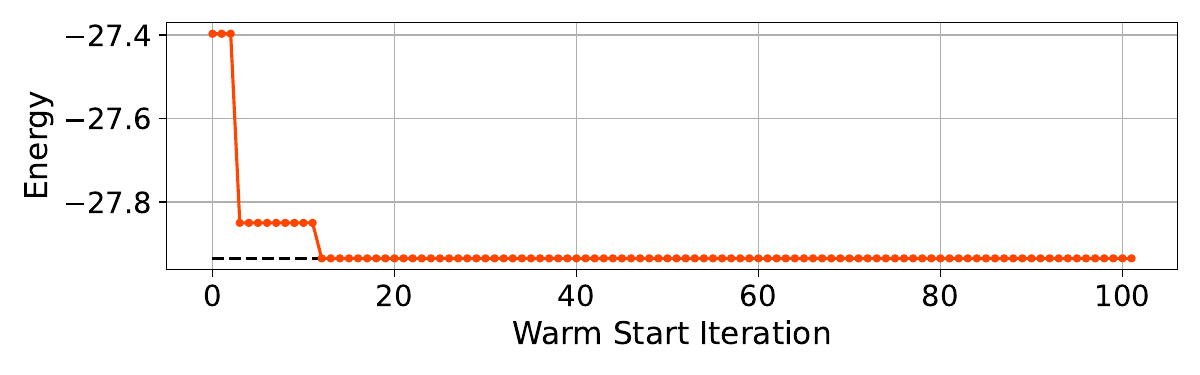}
    \includegraphics[width=0.35\textwidth]{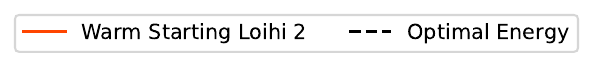}
    \caption{Iterated warm starting on results on Loihi~2 for $100$ iterations for each of the $16$ QUBO problems (each QUBO contains $64$ variables). Each panel shows the results for one of the $16$ QUBO problems. The minimum energy sample from the converged simulation is plotted at each iteration step. }
    \label{fig:iterated_warm_start_Loihi2}
\end{figure}

\subsection{Simulated Annealing Implementation}
\label{section:methods_simulated_annealing}

Simulated annealing is a classical heuristic algorithm which samples combinatorial optimization problems quite effectively \cite{kirkpatrick1983optimization}. The QUBO models described in Section~\ref{section:methods_transformation} can directly be sampled using a standard simulated annealing implementation, in this case we use the \texttt{dwave-neal} simulated annealing python software \footnote{\url{https://github.com/dwavesystems/dwave-neal}} in order to provide a reasonable basis of comparison between the evaluated algorithms and hardware.

Specifically, we use $1000$ reads per QUBO, while all other settings are set to their default values. This means that the annealing schedule is set to \texttt{geometric}. This implementation is a fast C++ implementation where a complete sweep of variable updates (Metropolis-Hastings updates) are performed in a fixed order for each temperature in the defined schedule of $\beta$ values. The range of $\beta$ values is computed based on the coefficients of the given QUBO problem \footnote{\url{https://github.com/dwavesystems/dwave-neal/blob/master/neal/sampler.py}}. The parameter $\beta$ denotes the inverse temperature $\frac{1}{k_BT}$ of the Boltzmann distribution, where $k_B$ is the Boltzmann constant and $T$ is the thermodynamic temperature \cite{kirkpatrick1983optimization}. In the default settings that we use, $1000$ Metropolis-Hastings updates are performed for each sample (where each update uses a different $\beta$ defined by the schedule, and each $\beta$ is utilized only once).

\subsection{Exact QUBO Solutions Using CPLEX}
\label{section:methods_cplex}
In order to determine how well all of the evaluated heuristic methods worked at sampling the optimal solution(s) of the generated QUBO models, we used the classical CPLEX \cite{CPLEX} solver which can solve the generated QUBOs exactly. CPLEX is a commercial mathematical optimization software. To this end, we first transform each QUBO into a Mixed Integer Quadratic Programming (MIQP) problem, and solve those problems exactly using the CPLEX solver \cite{CPLEX}. CPLEX deterministically finds and then verifies the optimal solution of a combinatorial optimization problem. Software such as CPLEX are used in the context of either generating good heuristic solutions within a compute time budget, or using as much compute time as is required to verify optimality of a solution -- in our case we use the later (i.e., no time limit) in order to compute the optimal energy (ground state energy) for these QUBO problems. CPLEX, although not as performant as new optimization software such as Gurobi, is a very standard classical benchmark for (general) classical optimization techniques. The CPLEX solver does not provide information on whether the QUBOs have multiple optimal solutions, or only $1$, however it is expected that in general these QUBOs will not have degenerate ground states. We leave further investigations on the ground state degeneracy of these binary sparse coding QUBO problems to future research.

%%%%%%%%%%%%%%%%%%%%%%%%%%%%%%%%%%%%%%%%%%%%%%%%%%%%%%%%%%%%%%
%%%%%%%%%%%%%%%%%%%%%%%%%%%%%%%%%%%%%%%%%%%%%%%%%%%%%%%%%%%%%%
%%%%%%%%%%%%%%%%%%%%%%%%%%%%%%%%%%%%%%%%%%%%%%%%%%%%%%%%%%%%%%
\section{Results}
\label{section:results}

Figure~\ref{figure:image_reconstructions} shows the combined best image reconstruction from sampling the $16$ QUBOs using the $4$ different methods. We observe that Loihi~2 provides a better first solution and has slight improvement with warm starting, while standard forward annealing is initially very poor, but achieves better average final solutions when reverse annealing QEMC is used.

Figure~\ref{fig:iterated_warm_start_Loihi2} presents the iterated warm started learning on Loihi~2 for all $16$ of the $64$ decision variable QUBO models. Interestingly, this process of warm starting the simulation using the state found during the previous computation seems to converge fairly fast to a local minimum for most of the QUBO problems. In these plots, each iteration is showing the minimum energy obtained from a set of $300$ samples obtained during the $6000$ time step simulation (at intervals of $20$ time steps). Notably, $4$ of the plots show high variability of the learning curve for those QUBOs, whereas for most of the other QUBOs there is very little variability of the learning curve. Figure~\ref{fig:iterated_warm_start_Loihi2} also plots the minimum energies computed using CPLEX. These results are the first that we are aware of for using iterated warm starting on a spiking neuromorphic processor.

\begin{figure}[htb!]
    \centering
    \includegraphics[width=0.24\textwidth]{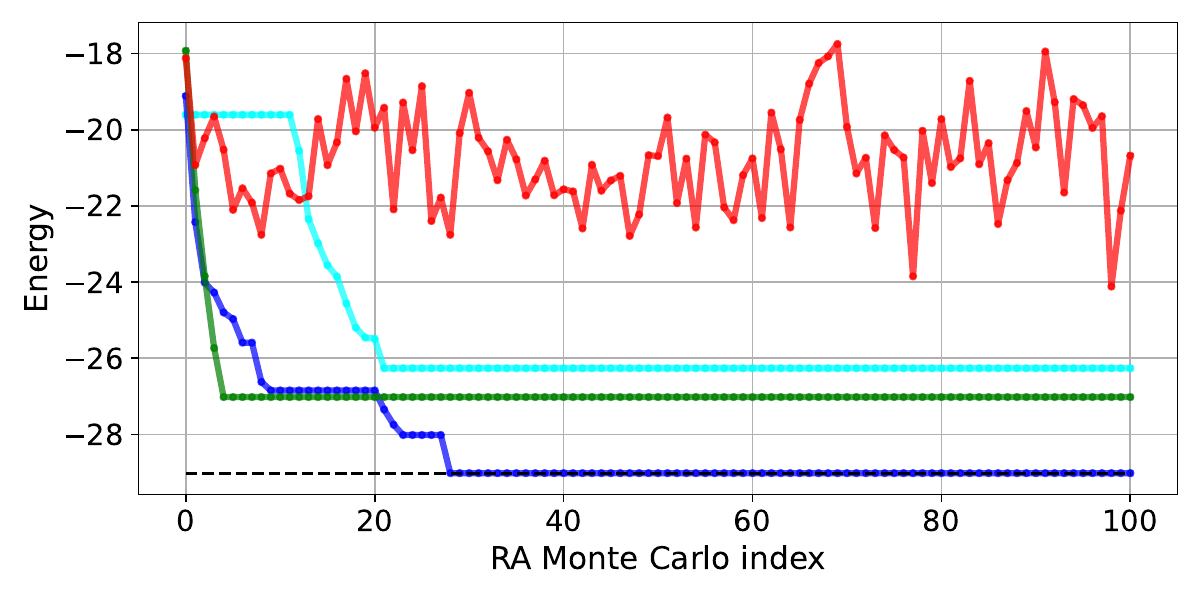}
    \includegraphics[width=0.24\textwidth]{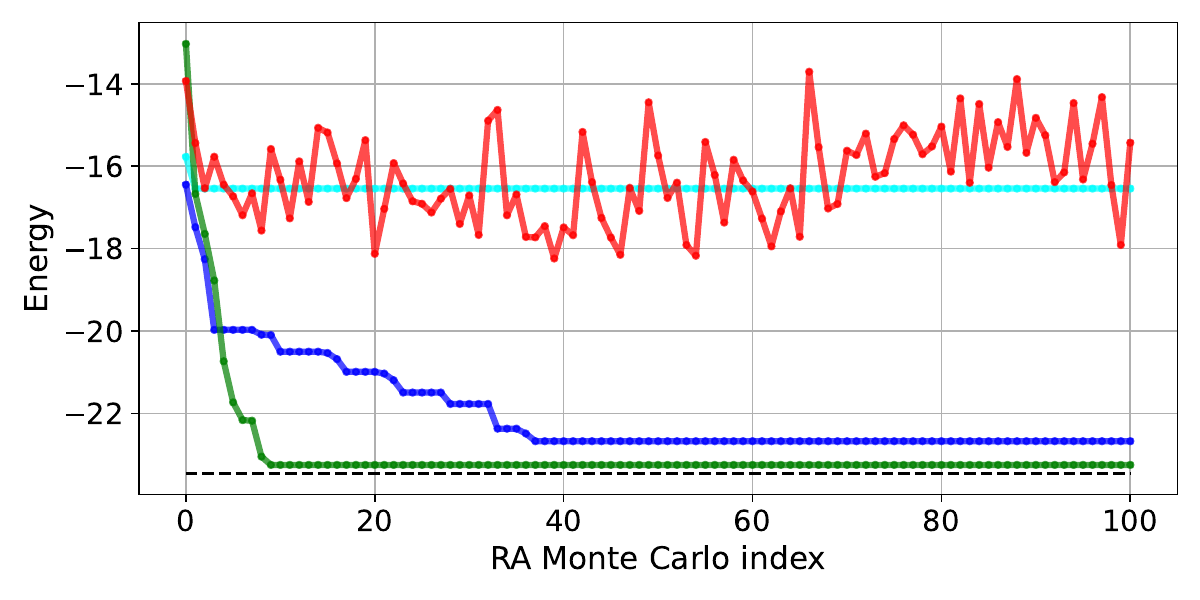}
    \includegraphics[width=0.24\textwidth]{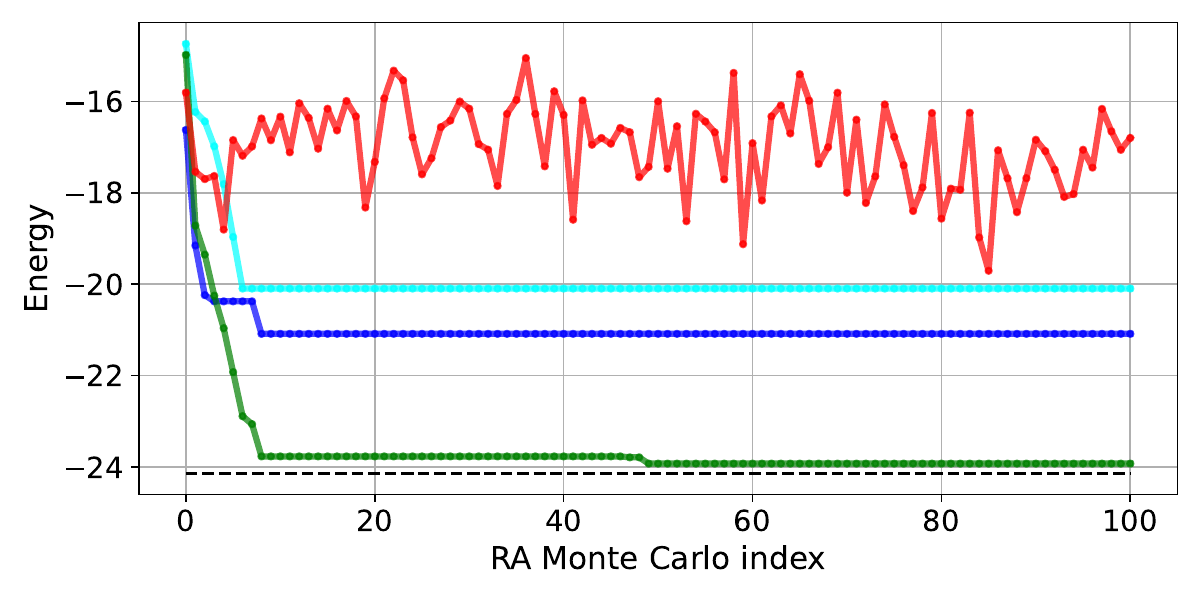}
    \includegraphics[width=0.24\textwidth]{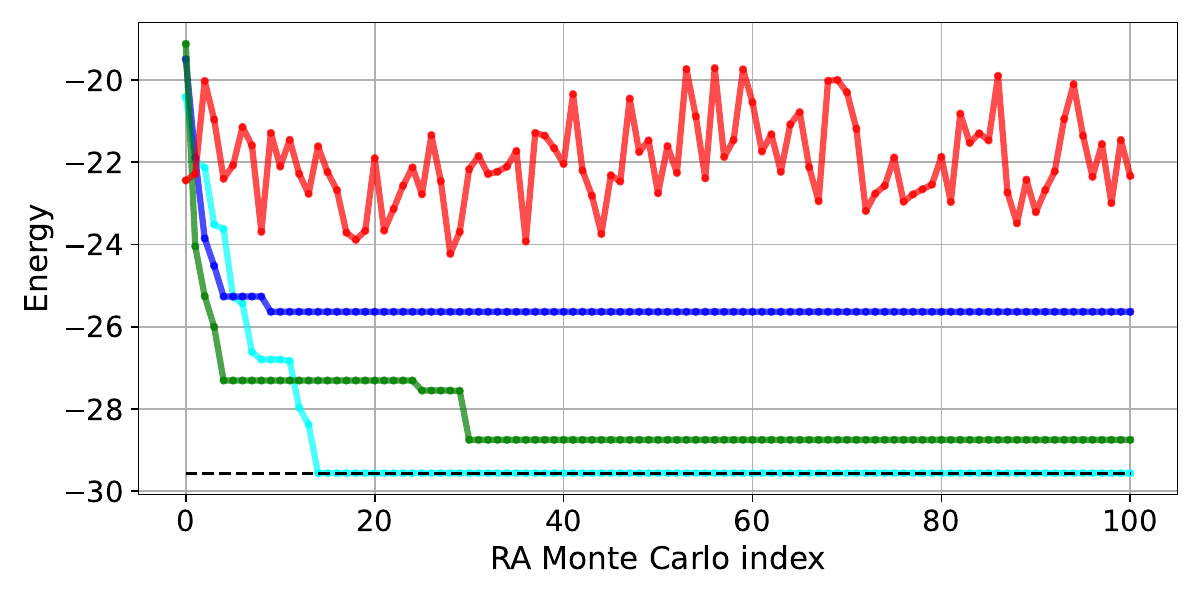}\\
    \includegraphics[width=0.24\textwidth]{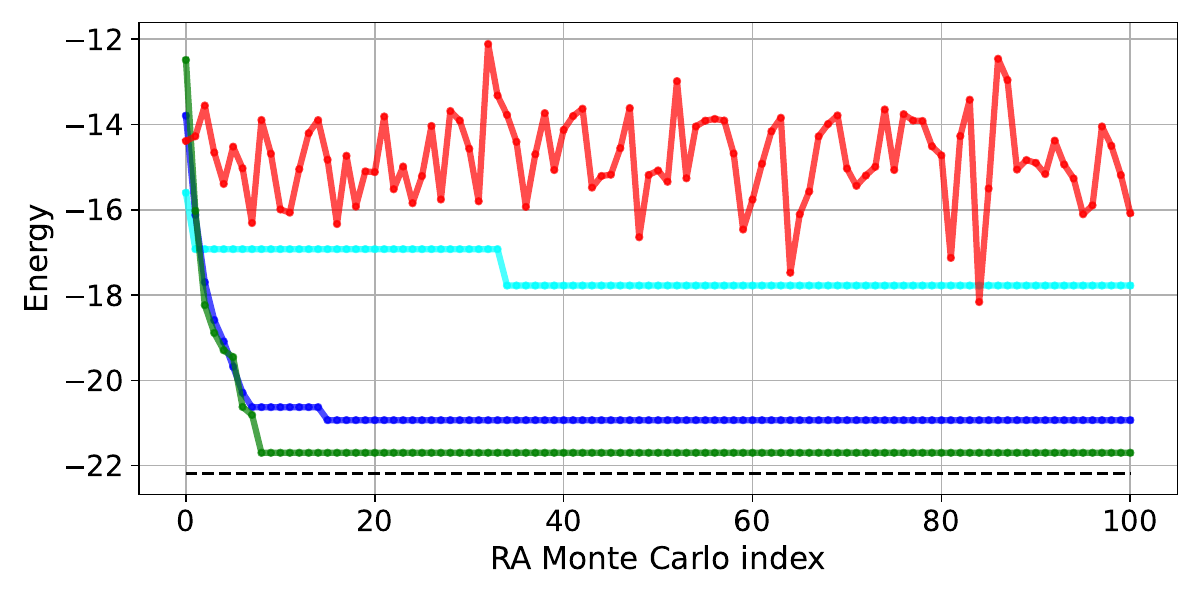}
    \includegraphics[width=0.24\textwidth]{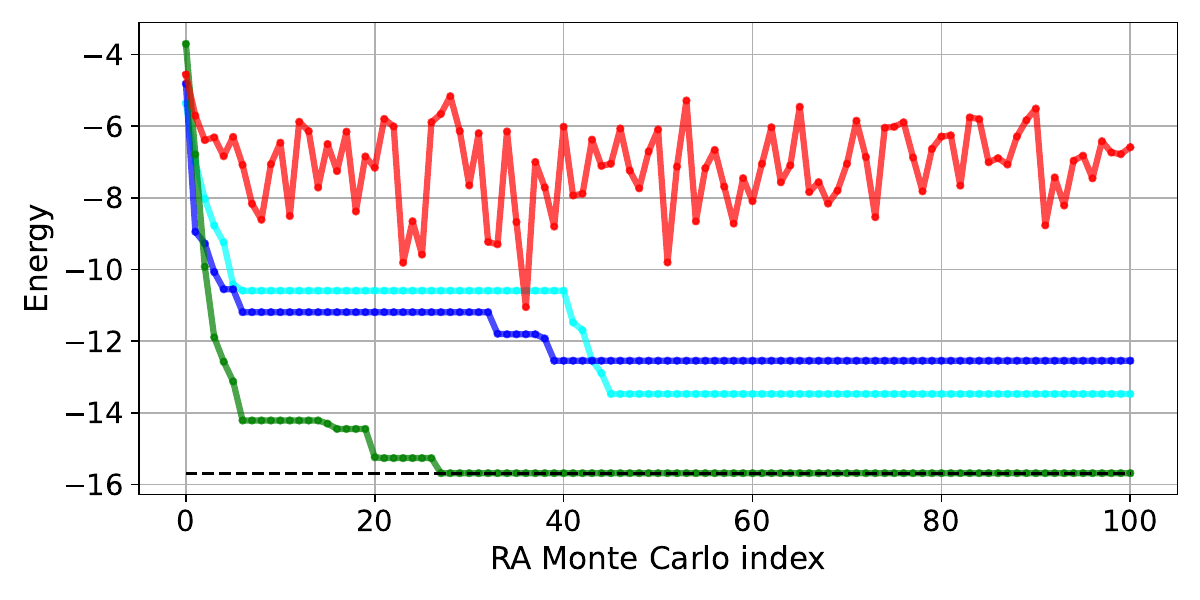}
    \includegraphics[width=0.24\textwidth]{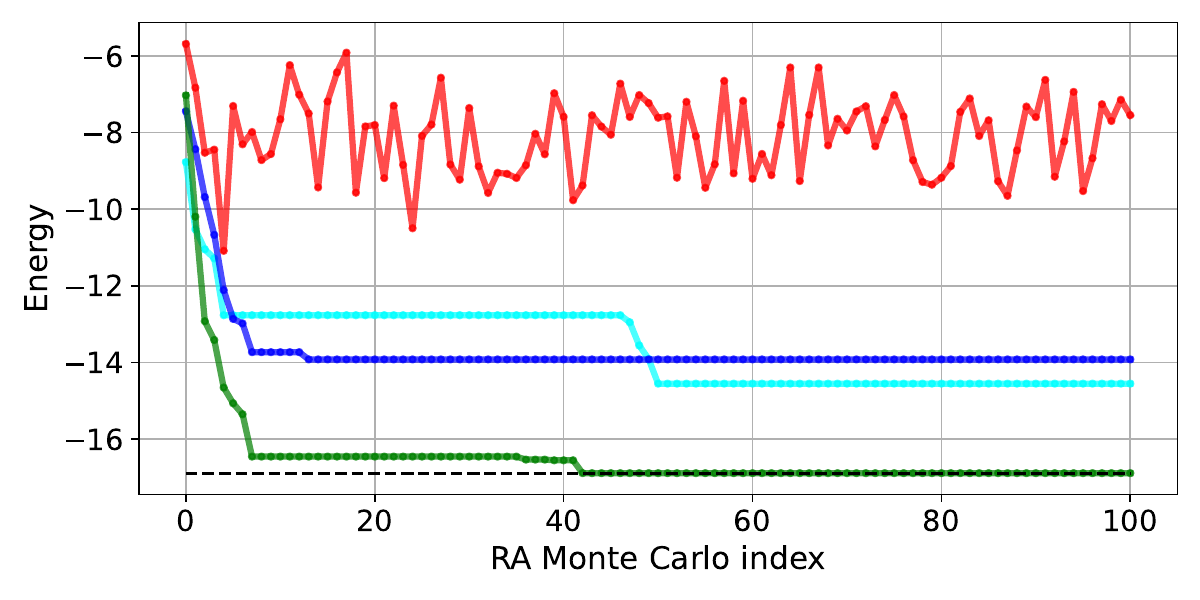}
    \includegraphics[width=0.24\textwidth]{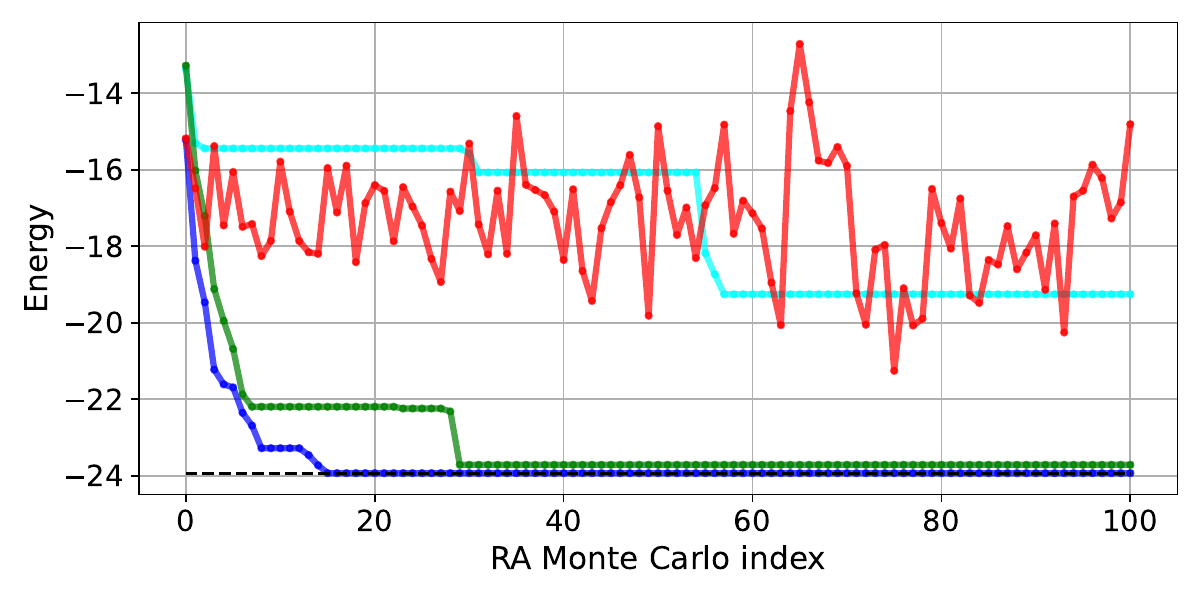}\\
    \includegraphics[width=0.24\textwidth]{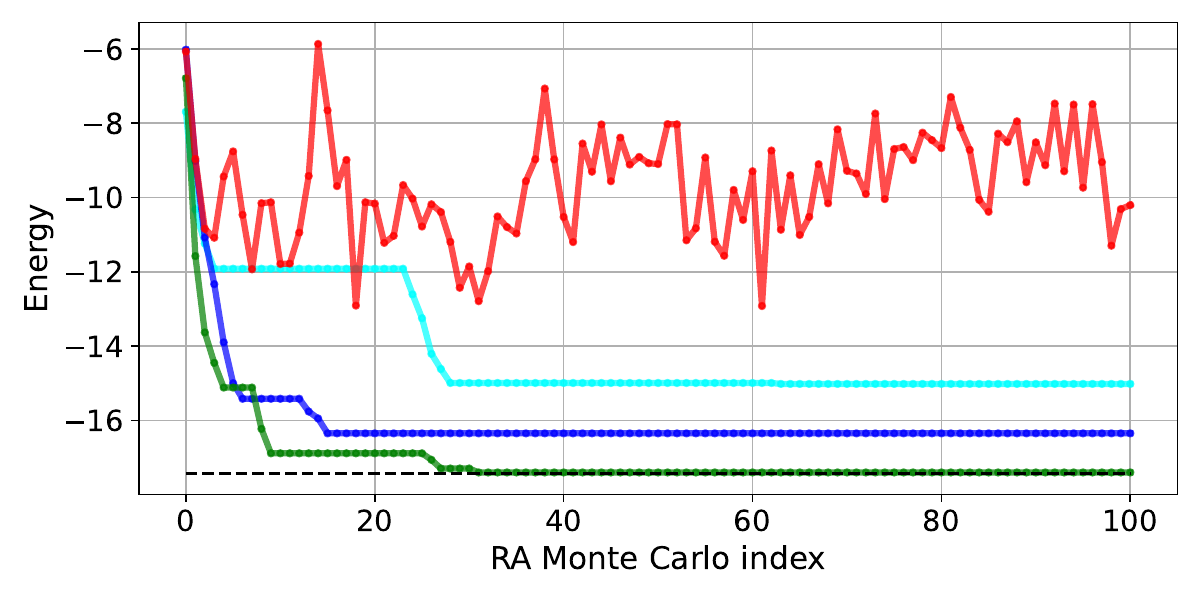}
    \includegraphics[width=0.24\textwidth]{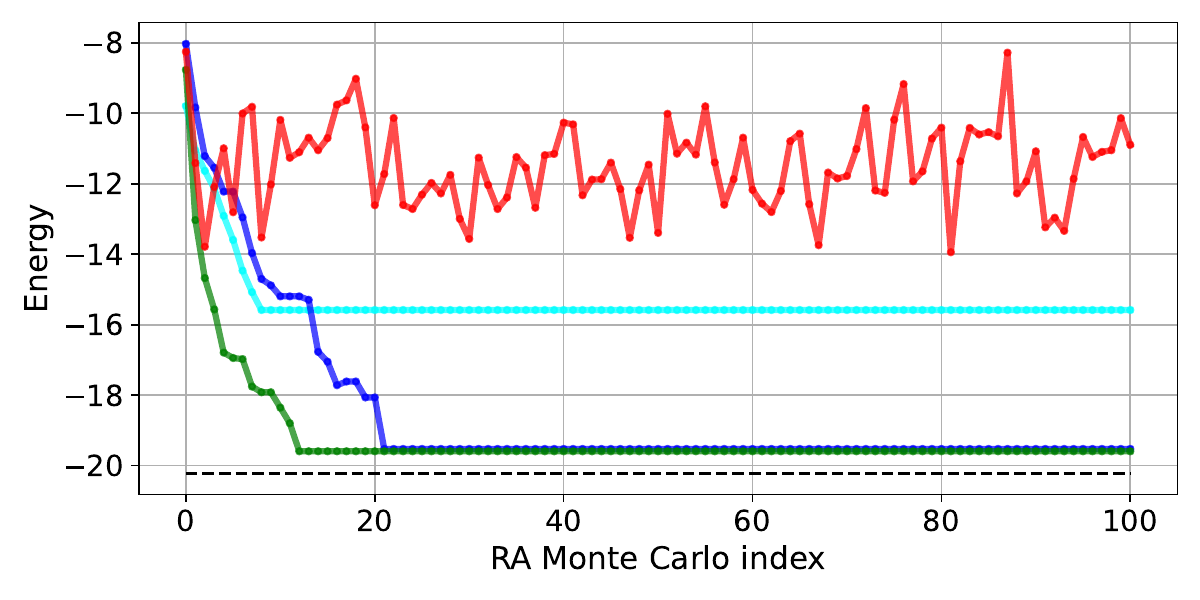}
    \includegraphics[width=0.24\textwidth]{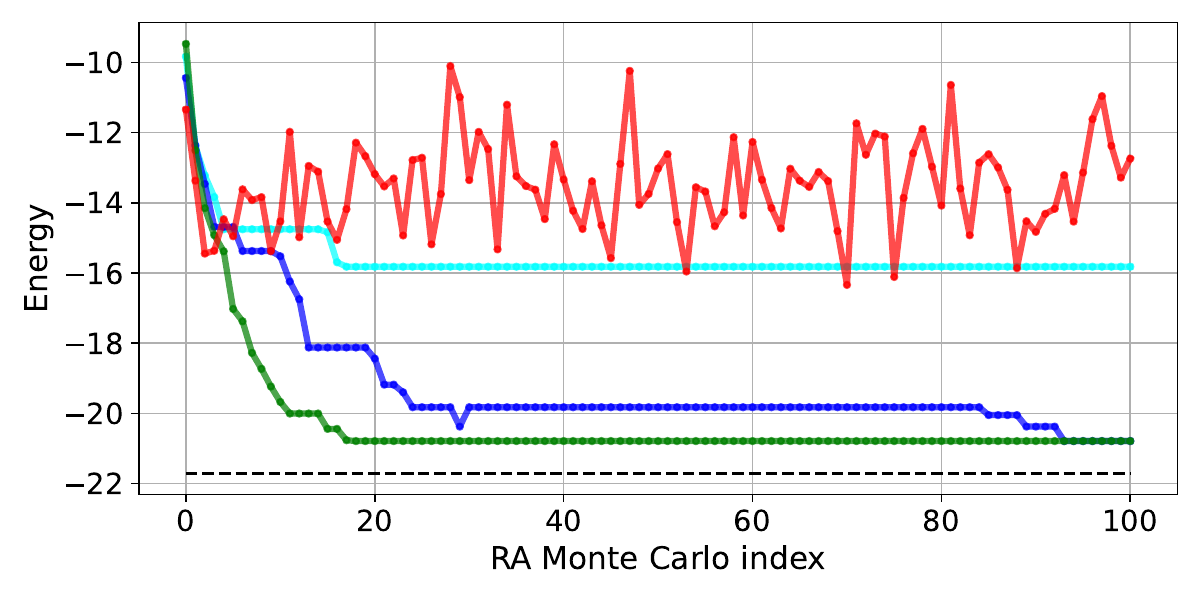}
    \includegraphics[width=0.24\textwidth]{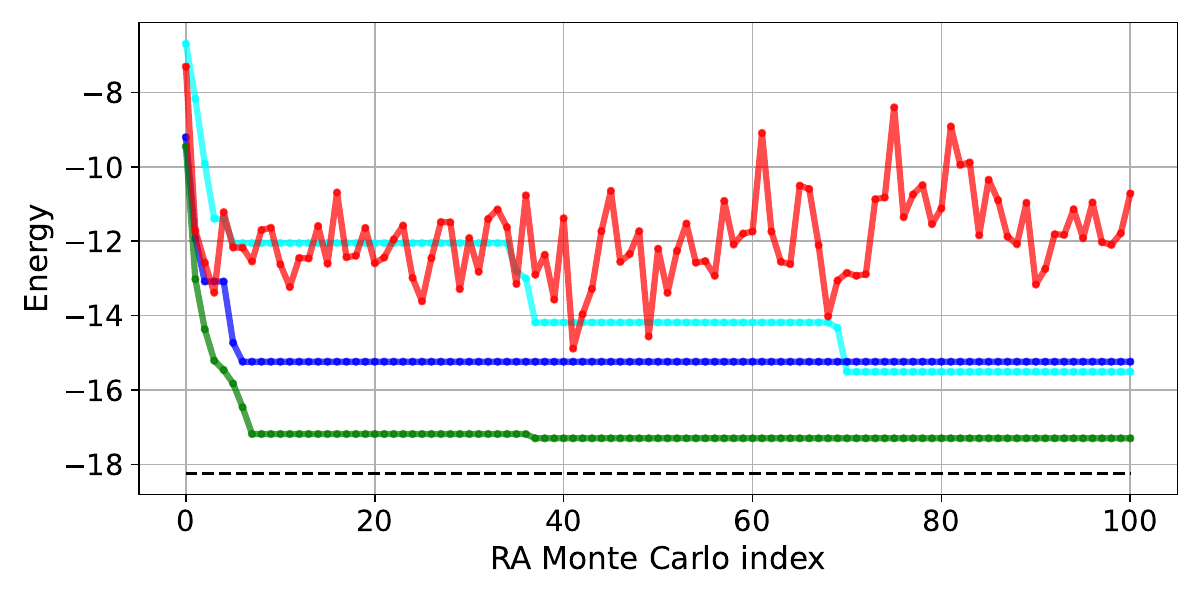}\\
    \includegraphics[width=0.24\textwidth]{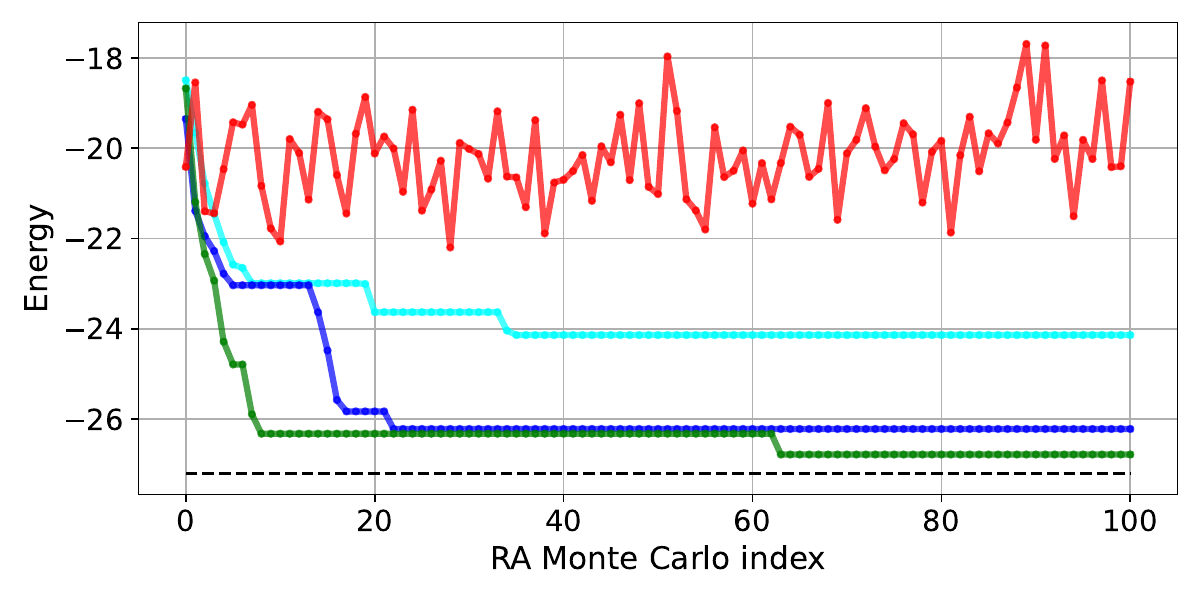}
    \includegraphics[width=0.24\textwidth]{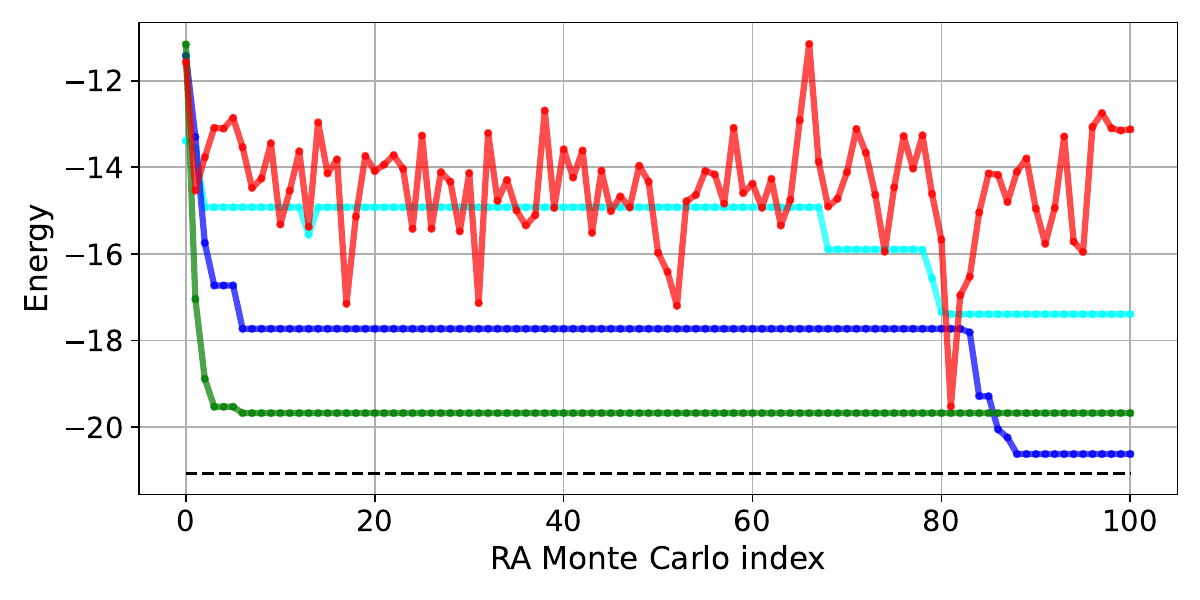}
    \includegraphics[width=0.24\textwidth]{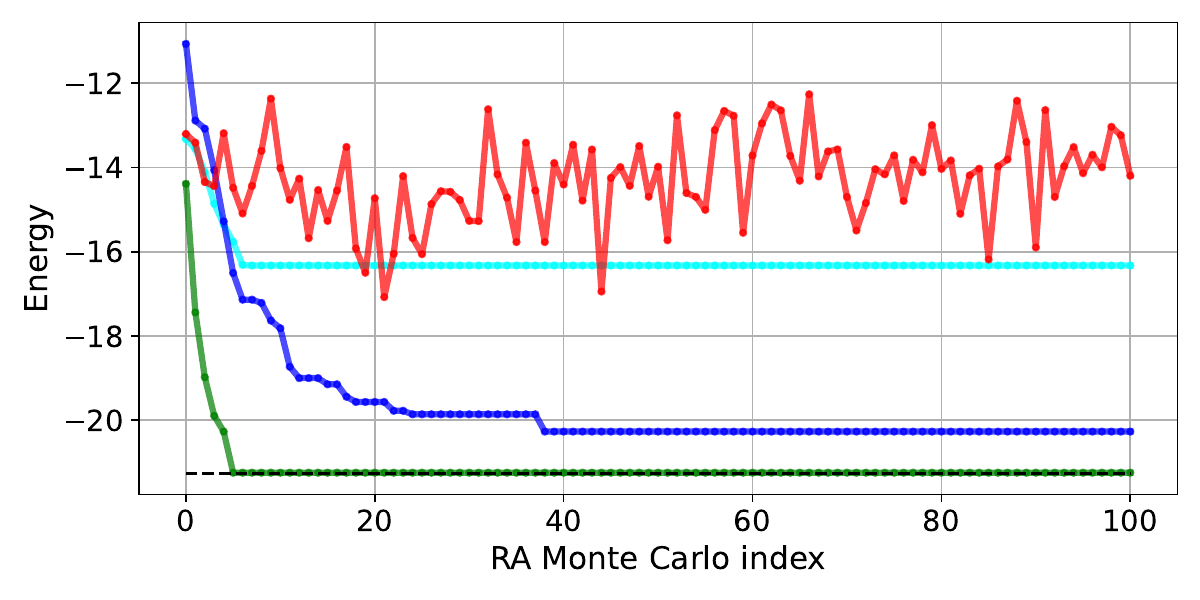}
    \includegraphics[width=0.24\textwidth]{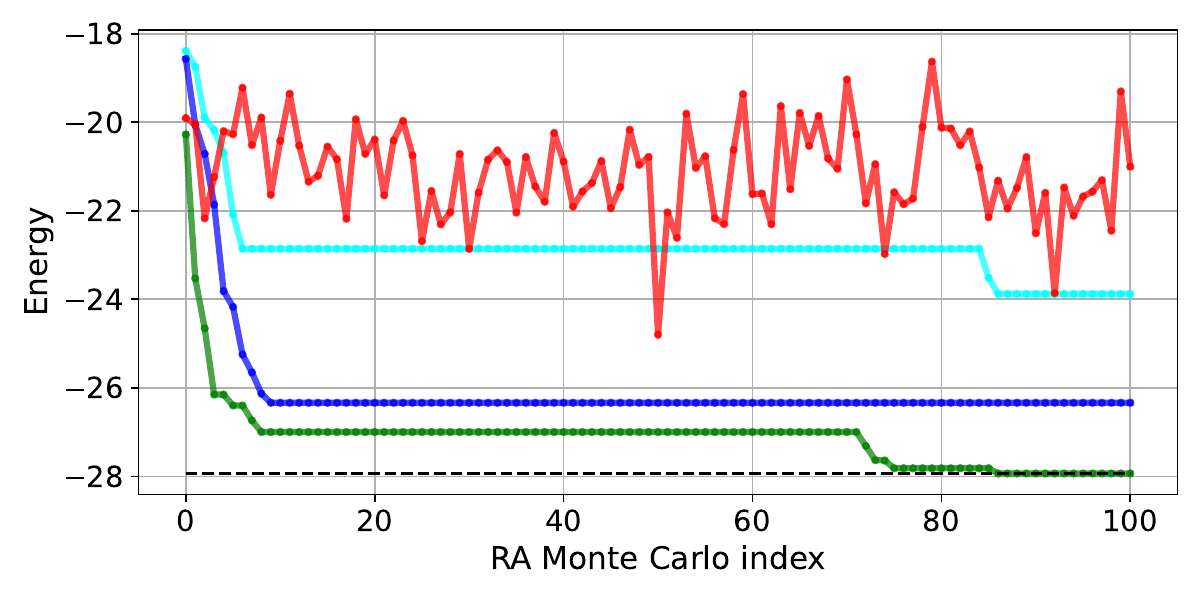}\\
    \includegraphics[width=0.49\textwidth]{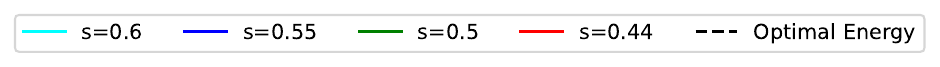}
    \caption{Quantum Evolution Monte Carlo with reverse annealing convergence. Each sub-plot shows the QEMC convergence for each of the $16$ separate, $64$-variable, QUBOs that are the set of patch sub-problems. The anneal fraction (and thereby the transverse field proportion relative to the diagonal problem Hamiltonian) at which the reverse anneal chain is paused is varied, and the resulting minimum energy found in the $1000$ samples at each Monte Carlo step is plotted. All quantum annealing computations used an annealing time of $100$ microseconds.}
    \label{fig:QEMC_64}
\end{figure}

Figure~\ref{fig:QEMC_64} shows the QEMC, or iterated reverse annealing, progression when sampling all of the $16$ sparse coding QUBOs using $100$ iterations and varying anneal fractions at which the pause occurred. At each QEMC step, the minimum energy (i.e., the objective function evaluation) sample is plotted. Figure~\ref{fig:QEMC_64} shows that there are some anneal fractions ($s$) that produced the desired monotonic decrease of minimum computed energy at each QEMC step, whereas other anneal fractions did not consistently improve on solution quality. Specifically, the anneal fraction of $s=0.44$ clearly introduced too much of the transverse field Hamiltonian into each simulation, resulting in a loss of information on the best previous obtained solution. On the other hand, the choice $s=0.6$ did not utilize enough of the quantum fluctuations provided by the transverse field Hamiltonian to search for lower energy states, although the improvement was always monotonic. The anneal fractions of $s=0.5$ and $0.55$ showed the best tradeoff in our experiments, allowing for reasonably good exploration of the problem search space. Figure~\ref{fig:QEMC_64} shows that the RA QEMC method was able to find the optimal solution for some, but not all of of the QUBO problems. This is examined in more detail in Table~\ref{table:minimum_energy_solutions}. Figure~\ref{fig:QEMC_64} plots the minimum energy computed by simulated annealing as black horizontal dashed lines, and Table~\ref{table:minimum_energy_solutions} shows that those solutions are also optimal as verified by CPLEX. When executing the QEMC simulations of Figure~\ref{fig:QEMC_64}, the initial state was chosen to be the best sample found from an initial standard linear schedule forward anneal; that initialization is then repeated with independent anneals for each run of QEMC, meaning that the initial states all have slightly different objective values. If there were multiple solutions with the same objective function cost, one was chosen arbitrarily as the initial vector. 

The ordering of the sub-plots (in terms of the QUBO problem being solved) in Figure~\ref{fig:iterated_warm_start_Loihi2} and Figure~\ref{fig:QEMC_64} is consistent between the two figures, thus facilitating a comparison. Notably, all of the iterated reverse annealing plots of Figure~\ref{fig:QEMC_64} show very clear and consistent convergence behavior (e.g., for $s=0.5$), but the iterated warm starting on Loihi~2 does not have as clear convergence behavior as a function of the iteration index. 

Table~\ref{table:minimum_energy_solutions} also reports the sparsity of the optimal bitstring computed by CPLEX. Simulated annealing always correctly found the optimal solution. Loihi~1 found the optimal solution for $0$ out of the $16$ QUBOs, so the optimal solution count is not shown. The Loihi~1 results are shown for comparison against Loihi~2, and are taken from the simulations in \cite{henke2023sampling}, where the minimum energies are taken from the distribution of $2000$ samples computed across a varied set of device parameters. The minimum energies reported both from iterated reverse quantum annealing and iterated warm starting on Loihi~2 are the minimum energies found from the entirety of the simulation (which, for converged simulations is the last point found after the $100$ iterations). The reported minimum energies from Loihi~2 are taken from the distribution of $100$ samples obtained from the iterated warm starting procedure. 

Table~\ref{table:minimum_energy_solutions} verifies that the trained sparse coding QUBOs indeed have solution vectors that are reasonably sparse. However, although unlikely to occur for our types of QUBO models, there could exist multiple variable assignments that give the same optimal objective function value (i.e., degenerate ground states). Moreover, as CPLEX only returns one optimal solution, the reported CPLEX results do not capture the case of more than one optimal solution with different sparsity. We do not report compute time in Table~\ref{table:minimum_energy_solutions} because it is relatively stable across different problem instances: the mean QPU time to obtain $1000$ samples on the D-Wave quantum processor is $1.3$ seconds, the mean wall-clock time to obtain $300$ samples on the Loihi~2 processor (this is $300$ samples taken during a single simulation sweep where we are measuring every $20$ time steps within the total simulation duration of $6000$ time steps) is $6.7$ seconds for the randomly initialized first iteration, and is $5.4$ seconds for each subsequent warm started iteration.

\begin{table*}[htb!]
\begin{center}
\begin{tabular}{ |p{1.0cm}||p{1.4cm}|p{1.4cm}|p{1.4cm}|p{1.4cm}|p{1.4cm}|p{2.2cm}|p{1.4cm}|p{1.4cm}| }
 \hline
 QUBO index & Ground state energy (CPLEX) & Optimal solution sparsity (CPLEX) & SA min. energy & SA min. energy count & QA QEMC min. energy & QA QEMC min. energy count & Loihi~1 min. energy & Loihi~2 iterated warm starting min. energy \\ 
 \hline
 \hline
 0 & $-29.016$ & $6 / 64$ & $-29.016$ & $826 / 1000$ & $-29.016$ & $394186/400000$ & $-26.308$ & $-28.0741$ \\ 
 \hline
 1 & $-23.4522$ & $12 / 64$ & $-23.4522$ & $25 / 1000$ & $-23.2522$ & $47038/400000$ & $-10.624$ & $-23.2522$ \\ 
 \hline
 2 & $-24.1331$ & $9 / 64$ & $-24.1331$ & $44 / 1000$ & $-23.9252$ & $20165/400000$ & $-15.784$ & $-23.4876$ \\ 
 \hline
 3 & $-29.5586$ & $8 / 64$ & $-29.5586$ & $117 / 1000$ & $-29.5586$ & $572222/400000$ & $-29.206$ & $-29.4979$ \\ 
 \hline
 4 & $-22.1755$ & $10 / 64$ & $-22.1755$ & $95 / 1000$ & $-21.6989$ & $14738/400000$ & $-15.369$ & $-21.8805$ \\ 
 \hline
 5 & $-15.6869$ & $10 / 64$ & $-15.6869$ & $26 / 10000$ & $-15.6869$ & $14816/400000$ & $-13.188$ & $-14.9114$ \\ 
 \hline
 6 & $-16.8891$ & $9 / 64$ & $-16.8891$ & $123 / 1000$ & $-16.8891$ & $10487/400000$ & $-12.744$ & $-15.8988$ \\ 
 \hline
 7 & $-23.9319$ & $9 / 64$ & $-23.9319$ & $103 / 1000$ & $-23.9319$ & $342411/400000$ & $-15.602$ & $-23.6264$ \\ 
 \hline
 8 & $-17.416$ & $10 / 64$ & $-17.416$ & $19 / 1000$ & $-17.3972$ & $11835/400000$ & $-13.432$ & $-16.7931$ \\ 
 \hline
 9 & $-20.2117$ & $7 / 64$ & $-20.2117$ & $45 / 1000$ & $-19.591$ & $16110/400000$ & $-16.036$ & $-18.6999$ \\ 
 \hline
 10 & $-21.703$ & $7 / 64$ & $-21.703$ & $41 / 1000$ & $-20.7899$ & $49428/400000$ & $-18.182$ & $-20.2913$ \\ 
 \hline
 11 & $-18.2355$ & $11 / 64$ & $-18.2355$ & $13 / 1000$ & $-17.2995$ & $13251/400000$ & $-8.9072$ & $-16.4587$ \\ 
 \hline
 12 & $-27.2012$ & $10 / 64$ & $-27.2012$ & $97 / 1000$ & $-26.7890$ & $24604/400000$ & $-21.610$ & $-26.621$ \\ 
 \hline
 13 & $-21.0627$ & $13 / 64$ & $-21.0627$ & $66 / 1000$ & $-20.6227$ & $59717/400000$ & $-8.747$ & $-20.2434$ \\ 
 \hline
 14 & $-21.254$ & $11 / 64$ & $-21.254$ & $29 / 1000$ & $-21.2437$ & $90453/400000$ & $-8.771$ & $-20.5983$ \\ 
 \hline
 15 & $-27.9347$ & $5 / 64$ & $-27.9347$ & $266 / 1000$ & $-27.9347$ & $14957/400000$ & $-26.593$ & $-27.9347$ \\ 
 \hline
 \hline
\end{tabular}
\end{center}
\caption{Optimal solutions and the best solutions found by simulated annealing, quantum annealing, and Loihi~2 for each of the $16$ QUBO problems (each QUBO contains $64$ variables). Objective function evaluations (e.g., energies) are rounded to a precision of 4 decimal places. The optimal solution sparsity is a fraction out of $64$, which shows how many of the variables are in a $1$ state. The RA QEMC sampling ground state energy success proportion is out of all samples (including parallel QA samples) across all parameter choices that were tested (specifically, the $4$ different anneal fractions at which the pause was set).}
\label{table:minimum_energy_solutions}
\end{table*}

%%%%%%%%%%%%%%%%%%%%%%%%%%%%%%%%%%%%%%%%%%%%%%%%%%%
%%%%%%%%%%%%%%%%%%%%%%%%%%%%%%%%%%%%%%%%%%%%%%%%%%%
%%%%%%%%%%%%%%%%%%%%%%%%%%%%%%%%%%%%%%%%%%%%%%%%%%%
\subsection{Binary Sparse Coding on 128 variables with QEMC}
\label{section:results_128_var_only}
\begin{figure}[htb!]
    \centering
    \includegraphics[width=0.49\textwidth]{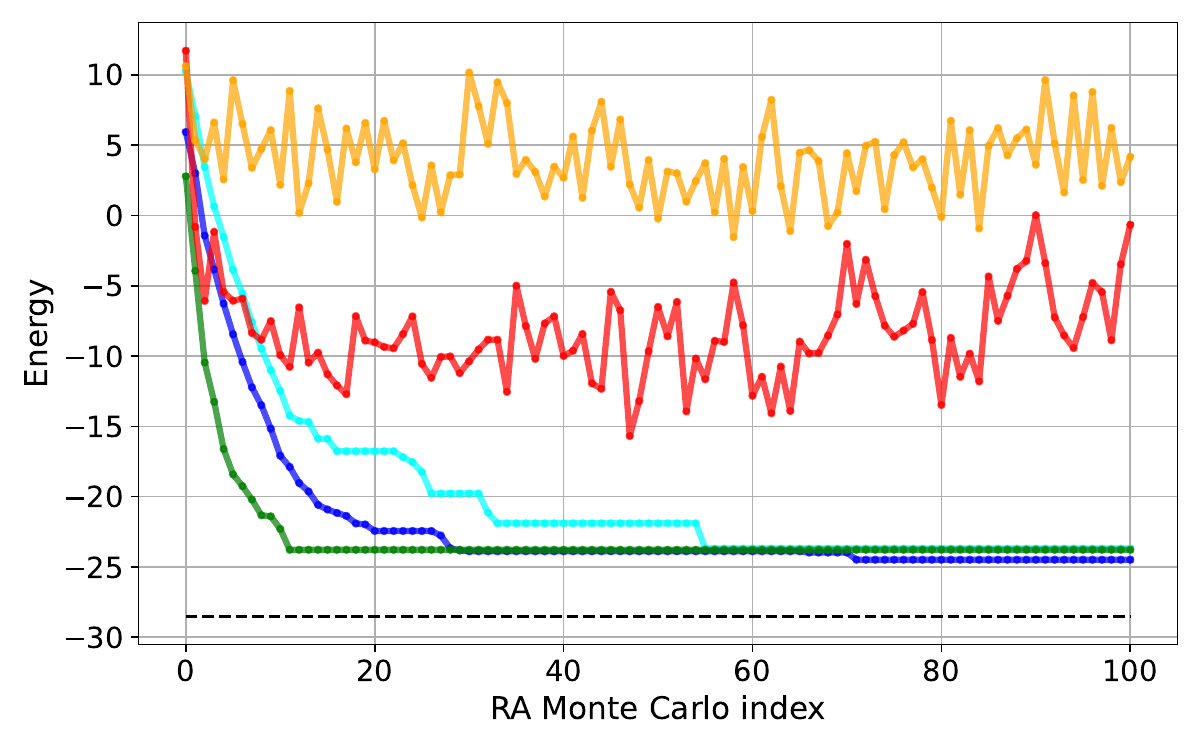}
    \includegraphics[width=0.49\textwidth]{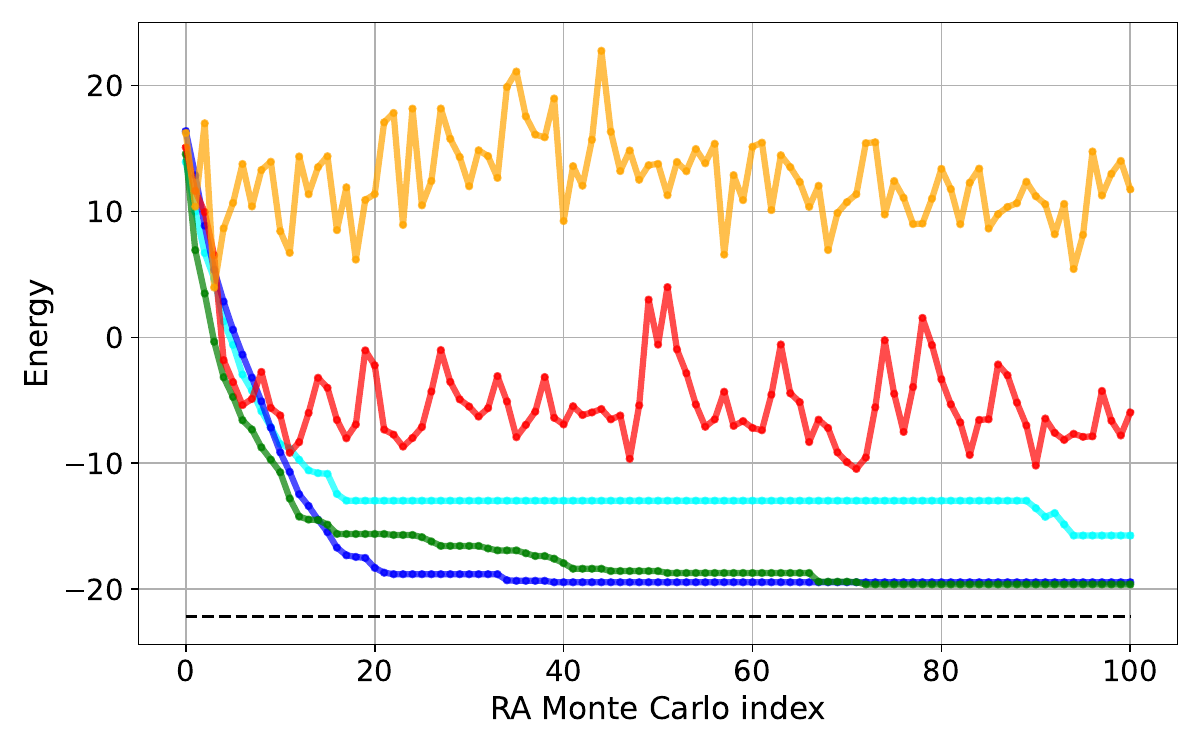}\\
    \includegraphics[width=0.49\textwidth]{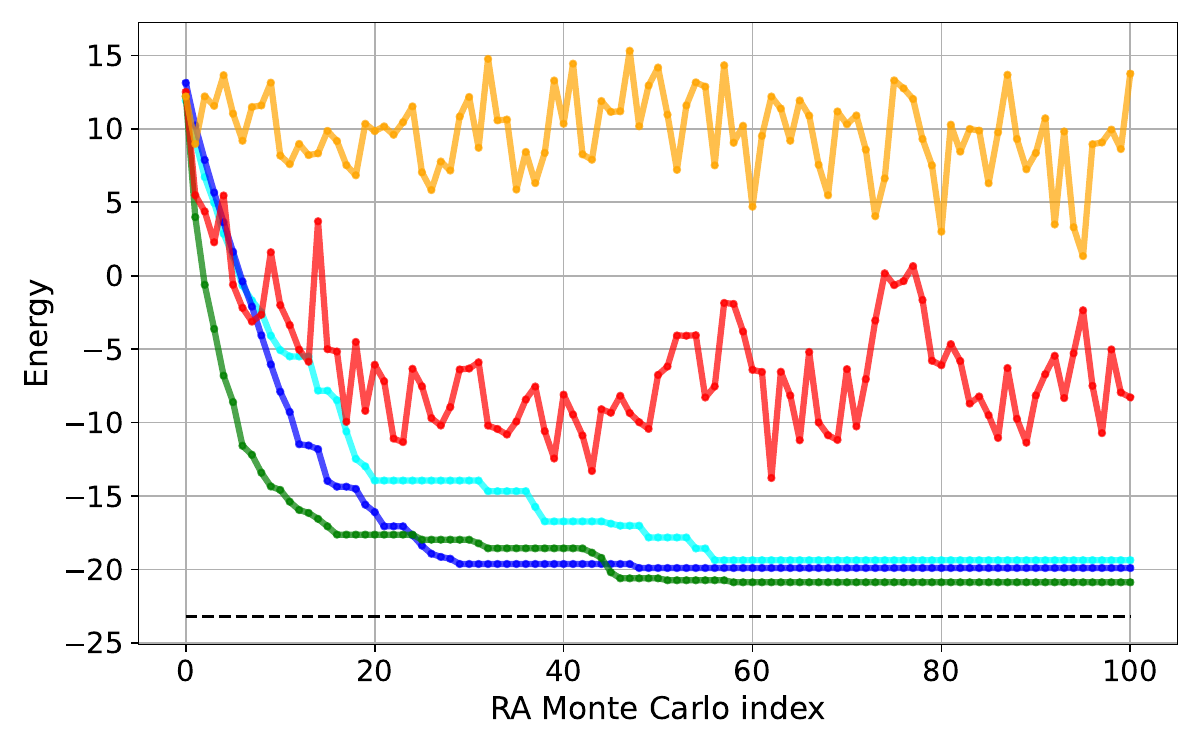}
    \includegraphics[width=0.49\textwidth]{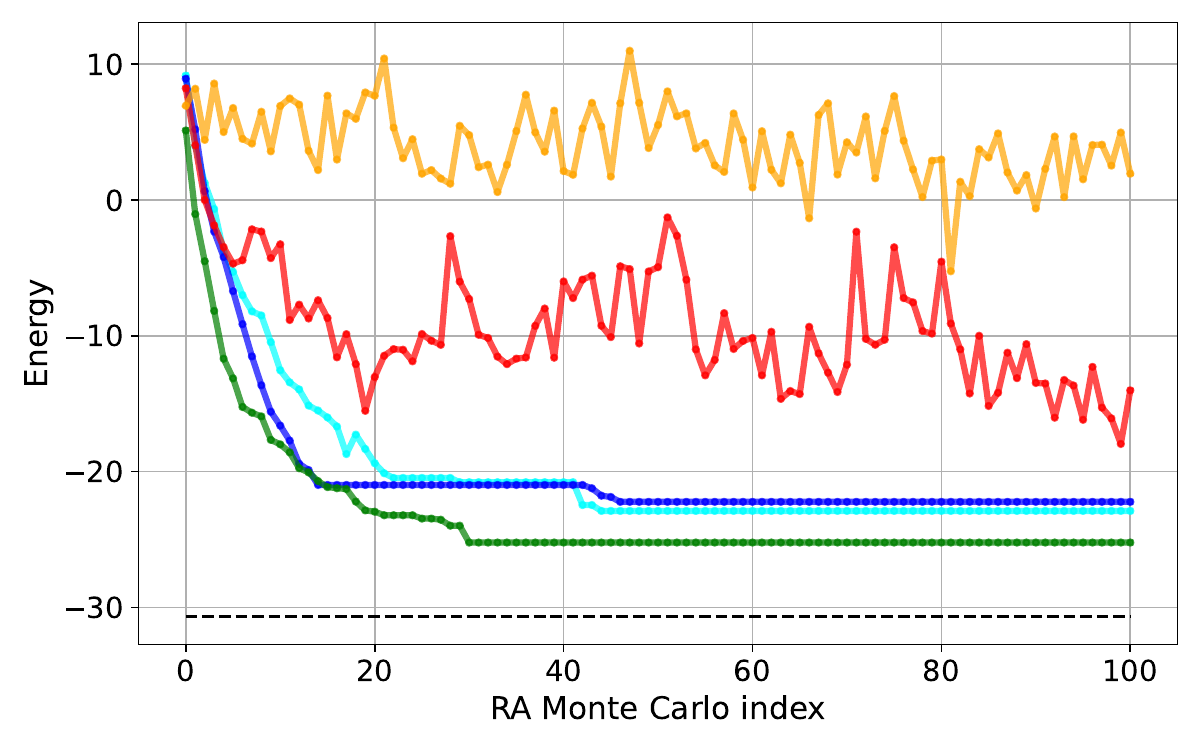}
    \includegraphics[width=0.68\textwidth]{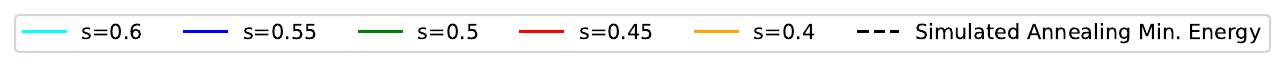}
    \caption{Quantum Evolution Monte Carlo with reverse annealing for 128 variable QUBOs. Each line is plotting the Monte Carlo simulation result minimum energy found (out of the $1000$ samples measured at each iteration) for different anneal fractions $s$ at which the anneal is (symmetrically) paused at during reverse annealing. }
    \label{fig:QEMC_128}
\end{figure}

Figure~\ref{fig:QEMC_128} shows the reverse annealing QEMC protocol applied to $128$ variable binary sparse coding QUBOs. These $128$ variable QUBOs are generated using the same fMNIST image as the $64$ variable QUBO partitioning reconstruction, but in this case the image is partitioned into $4$ QUBO models instead of $16$. The disjoint minor embeddings used for the $128$ RA QEMC procedure is shown in Figure~\ref{fig:embeddings_on_pegasus} (right), where only $2$ of the all-to-all random minor embeddings could be tiled. In Figure~\ref{fig:QEMC_128}, we see that QEMC fails to reach the solutions that simulated annealing finds. This is notable because it shows a very different performance compared to the $64$ variable QUBO problems. However, these results are consistent with the properties of minor embeddings with necessarily longer chains (in particular, resulting in higher chain break rates), especially considering the encoding precision requirements for these binary sparse coding QUBO problems. 

Both Figure \ref{fig:QEMC_64} and \ref{fig:QEMC_128} show that iterating reverse annealing, applied to these fully connected QUBOs, significantly improves the found solutions. In particular, when a good anneal fraction is used for the reverse anneal pause, the solution quality monotonically improves. This is is significant to see for such large minor embedded optimization problems. The effectiveness of iterated reverse annealing, or QEMC, has not been empirically demonstrated for minor embedded systems of this size, as far as we are aware. 

Like Figure \ref{fig:QEMC_64}, Figure~\ref{fig:QEMC_128} plots the best solution found when executing simulated annealing on the problem QUBOs; and like Figure \ref{fig:QEMC_64}, the simulated annealing solutions are also the optimal solutions found by CPLEX. In other words, the dashed horizontal black line in Figure~\ref{fig:QEMC_128} denotes the optimal energy of, or cost value, of these $128$ variable QUBOs. 

The best anneal fraction, which tradeoffs between exploration of new solutions and retaining memory of the initial spin state, is usually either $0.5$ or $0.55$ -- this is true for both Figure \ref{fig:QEMC_64} and Figure \ref{fig:QEMC_128}. Notably, using a reverse annealing pause at $s=0.4$ causes the Monte Carlo chain to not converge, but rather to oscillate approximately at the energy of the starting state at the beginning of the process.

%%%%%%%%%%%%%%%%%%%%%%%%%%%%%%%%%%%%%%%%%%%%%%%%%%%%%%%%%%%%%%
%%%%%%%%%%%%%%%%%%%%%%%%%%%%%%%%%%%%%%%%%%%%%%%%%%%%%%%%%%%%%%
%%%%%%%%%%%%%%%%%%%%%%%%%%%%%%%%%%%%%%%%%%%%%%%%%%%%%%%%%%%%%%
\section{Discussion}
\label{section:discussion}

In this article we have demonstrated that moderately sized binary sparse coding QUBO models can be sampled using both quantum annealing and neuromorphic computing technologies. These binary sparse coding QUBO problems were constructed using an un-supervised and un-normalized dictionary learning approach. We have found that the iterated reverse quantum annealing technique, implemented on a D-Wave quantum annealer, performed better than standard forward annealing. Similarly, iterated warm starting on Loihi~2 improved the solution quality of sampling the QUBOs. However, this energy improvement was marginal. We also found that the Loihi~2 sampling was better than the sampling done on Loihi~1 \cite{henke2023sampling}. Notably, a consistent trend for the Loihi processor(s) sampling is that the generated samples tend to be quite sparse -- more sparse than sub-optimal sampling from standard simulated annealing or quantum annealing. 

Amongst others, the following reasons might have impeded the forward annealing and reverse annealing sampling from reaching the global minimum in our experiments. The first, and almost certainly the most important, is the coefficient precision limits that can be programmed on current D-Wave devices. The sparse coding QUBOs have quadratic terms which are defined with high precision, and are nearly all very close to zero (see Figure~\ref{fig:QUBO_matrices}). It is very likely that such sparse coding QUBO coefficients require higher precision than what the D-Wave hardware supports. Note that the precision limits on the hardware are not a well defined bit precision, but rather are influenced by a number of factors on the analog hardware, such as asymmetric digital-to-analog converter quantization, making the exact precision limits hard to know for all problem types and simulation settings \footnote{\url{https://docs.dwavesys.com/docs/latest/c_qpu_ice.html}}. The next obstacle is the use of a minor embedding. A minor embedding is necessary for implementing general QUBO problems on D-Wave, however such embeddings have ferromagnetic chains which dominate the energy scale programmed on the chip, therefore not allowing a larger range of the logical problem coefficients to be effectively used in the computation. Additionally, although majority vote unembedding is used, chain breaks that are present in the samples represent likely sub-optimal solution quality. Analog control errors when the hardware is representing the diagonal Hamiltonian on the physical machine can also lead to errors in the simulation. Finally, interaction with the environment causes the system to decohere and noise to be introduced into the simulation.

For future research, it would be interesting to be able to formulate sparse coding QUBO problems that match a desired connectivity graph. This would be particularly useful for generating hardware-compatible QUBOs for D-Wave hardware graphs, as it would allow one to considerably scale up the size of QUBO problems that can be solved (up to the size of the entire hardware graph, which has over $5000$ qubits). In particular, probing the scaling of solution quality for future generations of both neuromorphic processors and quantum annealing processors would be of interest to compare against existing classical optimization algorithms. In general, creating significantly larger sparse coding QUBOs using un-normalized dictionary learning \cite{henke2023sampling}, and then evaluating methods to find good solutions to those larger QUBOs, would be of interest for future studies.

\appendix

\section{Parallel Minor Embeddings}
\label{section:appendix_parallel_minor_embeddings}

\begin{figure}[ht!]
    \centering
    \includegraphics[width=0.45\textwidth]{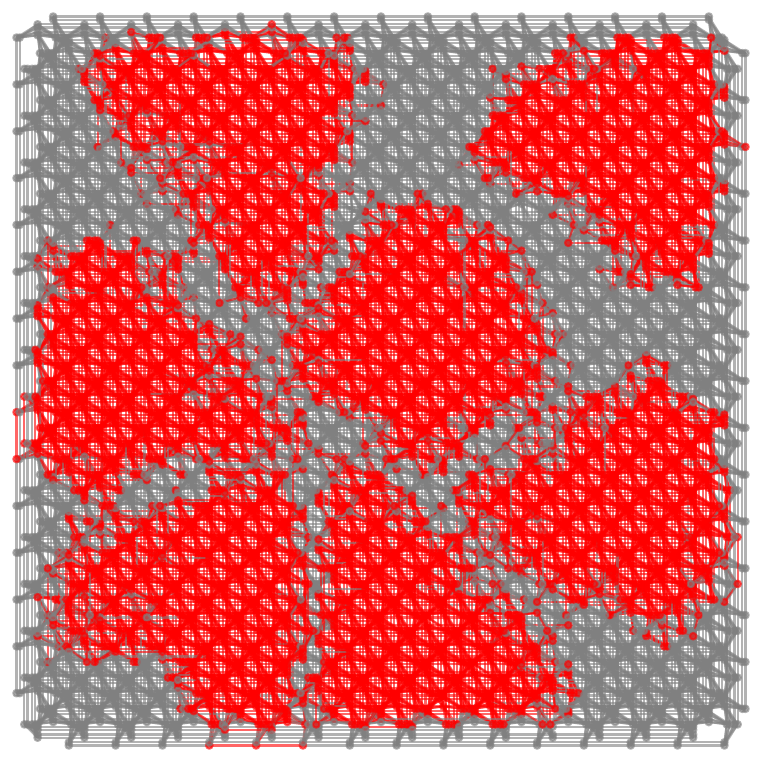}\hfill
    \includegraphics[width=0.45\textwidth]{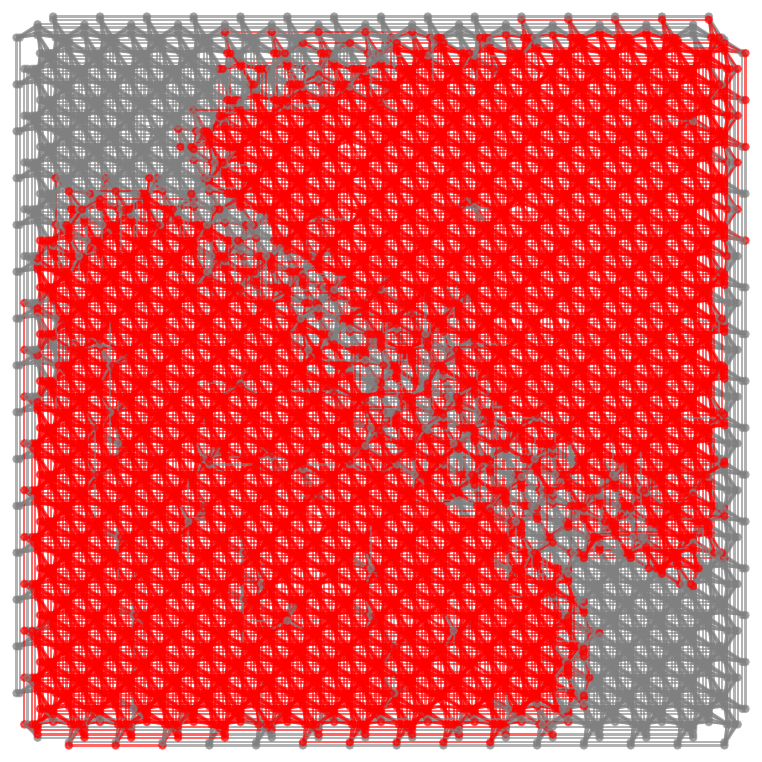}
    \caption{Parallel quantum annealing fully connected minor-emebddings on the hardware graph of \texttt{Advantage\_system4.1}. Seven disjoint minor embeddings of size $64$ variables with all-to-all connectivity (left), and two disjoint minor embeddings of size $128$ variables with all-to-all connectivity (right) on the Pegasus $P_{16}$ hardware graph of the \texttt{Advantage\_system4.1}.
    \label{fig:embeddings_on_pegasus}}
\end{figure}

Figure~\ref{fig:embeddings_on_pegasus} shows the parallel minor embeddings on the Pegasus hardware graph.

\begin{figure}[ht!]
    \centering
    \includegraphics[width=0.40\textwidth]{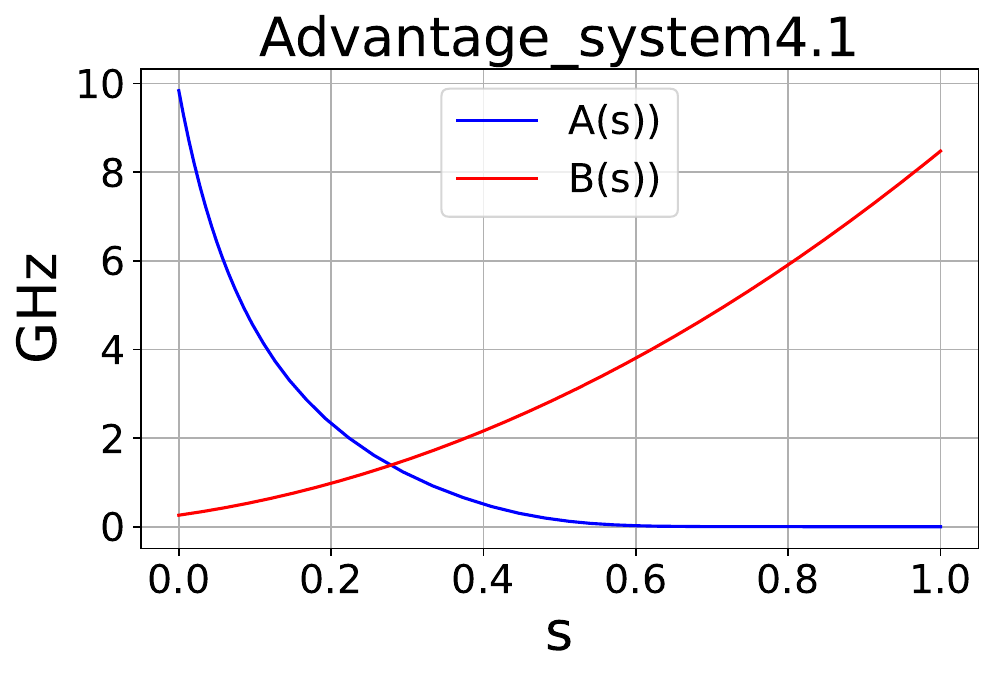}
    \caption{Calibrated anneal schedule for the D-Wave quantum annealing processor used in this study. The exact $A(s)$ and $B(s)$ values are plotted in units of GHz as a function of the anneal fraction $s \in [0,1]$.}
    \label{fig:DWave_schedule}
\end{figure}

\section{D-Wave Quantum Annealer Schedule Calibration}
\label{section:appendix_D_Wave_anneal_schedule_calibration}

Figure~\ref{fig:DWave_schedule} gives the exact vendor-provided anneal schedule calibration data. This data is publicly available on the D-Wave website under QPU-Specific Characteristics \footnote{\url{https://docs.dwavesys.com/docs/latest/doc_physical_properties.html}}.